\documentclass[tightenlines,twoside,secnumarabic,
               onecolumn,floatfix,nofootinbib,preprint,11pt]{revtex4-1}
\usepackage{graphicx}
\usepackage{dcolumn}
\usepackage{float}
\usepackage{amsmath}
\usepackage{amsfonts}
\usepackage{multirow}
\usepackage{bbold}
\usepackage{color}
\usepackage{hyperref}
\usepackage{subscript}

\begin{document}
\newcommand\unit{\mathbb{1}}
\def\Lt{\overline{L}}
\def\nn{{++}}
\def\an{{+-}}
\def\bn{{+\pm}}
\def\sl{{\rm sl}}
\def\ax{{\rm ax}}
\def\vect{{\rm v}}
\def\ib{{\; \bar\imath}}
\def\Atree{A^{\rm tree}}
\def\half{\textstyle \frac{1}{2}}
\def\nf{n_{\mskip-2mu f}}
\def\q{{q}}
\def\Q{{Q}}
\def\qb{{{\bar q}}}
\def\Qb{{{\bar Q}}}
\def\sl{{\rm sl}}
\def\ax{{\rm ax}}
\def\A#1{{\cal A}_{#1}}
\def\vect{{\rm v}}
\def\cM{\mathcal{M}}
\def\prop#1{{\cal P}_{#1}}
\def\Af{v_A^f}
\def\Vf{v_V^f}
\def\fudgeBDK{-}
\def\mfudgeBDK{\relax}
\def\tree{{\rm tree}}
\def\sutqm{(s_{13}+s_{14})}
\def\sdtqm{(s_{23}+s_{24})}
\def\sucsm{(s_{15}+s_{16})}
\def\sdcsm{(s_{25}+s_{26})}
\def\sutq{{s}_{134}}
\def\sdtq{{s}_{234}}
\def\sucs{{s}_{156}}
\def\sdcs{{s}_{256}}
\def\sud{{s}_{12}}
\def\stq{{s}_{34}}
\def\scs{{s}_{56}}
\def\sut{{s}_{13}}
\def\suq{{s}_{14}}
\def\sdt{{s}_{23}}
\def\sdq{{s}_{24}}
\def\suc{{s}_{15}}
\def\sus{{s}_{16}}
\def\sdc{{s}_{25}}
\def\sds{{s}_{26}}
\def\duxtqxd{1}
\def\ddxuxtq{2}
\def\duxdxtq{3}
\def\cuxd{1}
\def\cudxtq{2}
\def\cuxtq{3}
\def\cdxtq{4}
\def\cuxcs{5}
\def\cdxcs{6}
\def\cD{\cal D}
\def\cP{\cal P}
\def\cg{c_\Gamma}
\def\gW{g_W}
\def\rG{r_\Gamma}
\def\jb{i_{b}}
\def\jt{i_{t}}
\def\tb{\bar{t}}
\def\eb{\bar{e}}
\def\mub{\bar{\mu}}
\def\qb{\bar{q}}
\def\spaa#1.#2.#3{\langle\mskip-1mu{#1}|#2|{#3}\mskip-1mu\rangle}
\def\spbb#1.#2.#3{[\mskip-1mu{#1}|#2|{#3}\mskip-1mu]}
\def\spa#1.#2{\left\langle#1\,#2\right\rangle}
\def\spb#1.#2{\left[#1\,#2\right]}
\def\spab#1.#2.#3{\left\langle#1|#2|#3\right]}
\def\spba#1.#2.#3{\left[#1|#2|#3\right\rangle}
\def\spbab#1.#2.#3.#4{[\mskip-1mu{#1}
                  | #2  #3 | {#4}\mskip-1mu]}
\def\spaba#1.#2.#3.#4{\langle\mskip-1mu{#1}
                  | #2  #3 | {#4}\mskip-1mu\rangle}
\def\Wzb{\bar{W}_0}
\def\Pzb{\bar{P}_0}
\def\Pthreeb{\bar{P}_3}
\def\P3b{\bar{P}_3}
\def\Ypb{\bar{Y}_p}
\def\Ypz{{Y}_p(z)}
\def\Ywb{\bar{Y}_w}
\def\Ppb{\bar{P}_+}
\def\Pmb{\bar{P}_-}
\def\Wpb{\bar{W}_+}
\def\Wmb{\bar{W}_-}
\def\cO{{\cal O}}
\def\li{{\rm Li_2}}
\def\g0{\gamma_0}
\def\gp{\gamma^{+}}
\def\gm{\gamma^{-}}
\def\lp{\gamma^{+}}
\def\lm{\gamma^{-}}
\def\xp{x_{+}}
\def\xm{x_{-}}
\def\bentarrow{\:\raisebox{1.3ex}{\rlap{$\vert$}}\!\rightarrow}                 
\def\dkp#1#2#3#4{
\begin{array}{r c l}
#1 & \rightarrow & #2#3 \\
 & & \phantom{\; #2}\bentarrow #4
\end{array}}                                                                    
\def\bothdk#1#2#3#4#5{
\begin{array}{r c l}                                                            
#1 & \rightarrow & #2#3 \\
 & & \:\raisebox{1.3ex}{\rlap{$\vert$}}\raisebox{-0.5ex}{$\vert$} 
\phantom{#2}\!\bentarrow #4 \\
 & & \bentarrow #5                                                              
\end{array}                                                                     
}                                                                               
\newcommand{\kirill}{\colour{red}}
\newcommand{\comment}[1]{{\bf [#1]}}
\newcommand{\beq}{\begin{equation}}
\newcommand{\eeq}{\end{equation}}
\newcommand{\beqn}{\begin{eqnarray}}
\newcommand{\eeqn}{\end{eqnarray}}
\newcommand{\bi}[1]{\bibitem{#1}}
\newcommand{\fr}[2]{\frac{#1}{#2}}
\newcommand{\non}{\nonumber}
\newcommand{\Et}{E_t}
\newcommand{\Pt}{P_t}
\newcommand{\pt}{p_t}
\newcommand{\pb}{p_b}
\newcommand{\pw}{p_W}
\newcommand{\pg}{p_g}
\newcommand\tpW        {{\tilde p}_W}
\newcommand\tpb        {\tilde p_b}
\newcommand{\ar}{\mbox{$\rightarrow$}}
\def\ra{\rightarrow}

\newcommand{\slsh}{\rlap{$\;\!\!\not$}}     
\def\amuh{a_\mu^{{\mathrm had}}}
\def\vec#1{{\mbox{\boldmath$#1$}}}
\def\ket#1{\vert #1 \rangle}
\def\bra#1{\langle #1 \vert}
\newcommand{\as}{\alpha_S}
\newcommand{\p}{\mbox{$\vec{p}$}}
\newcommand{\pp}{\mbox{$\vec{p}'$}}
\newcommand{\rp}{\mbox{$\vec{r}'$}}
\newcommand{\kp}{\mbox{$\vec{k}'$}}
\newcommand{\e}{\mbox{$\vec{e}$}}
\newcommand{\s}{\mbox{$\vec{s}$}}
\newcommand{\Li}{{\rm Li}}
\newcommand{\lsim}{\mbox{\raisebox{-0.3ex}{%
\footnotesize $\:\stackrel{<}{\sim}\:$}} }
\newcommand{\gsim}{\mbox{\raisebox{-0.3ex}{%
\footnotesize $\:\stackrel{>}{\sim}\:$}} }
\newcommand{\lb}{\left (}
\newcommand{\rb}{\right )}
\newcommand{\ep}{\epsilon}
\newcommand{\vep}{\epsilon}
\newcommand{\dd}{{\rm d}}
\newcommand{\om}{\omega}

\newcommand{\sS}{\mbox{$\vec{\sigma}\vec{\sigma}'$}}
\newcommand{\si}{\mbox{$\vec{\sigma}$}}
\newcommand{\vgamma}{\mbox{$\vec{\gamma}$}}
\newcommand{\vxi}{\mbox{$\vec{\xi}$}}

\newcommand{\pop}[1]{\mbox{$\Lambda_+(#1)$}}
\newcommand{\nep}[1]{\mbox{$\Lambda_-(#1)$}}
\newcommand{\Dafne}{DA$\Phi$NE}
\newcommand{\mar}{\marginpar{***}}
\def\spab#1.#2.#3{\langle\mskip-1mu{#1}
                  | #2 | {#3}\mskip-1mu]}
\def\spba#1.#2.#3{[\mskip-1mu{#1}
                  | #2 | {#3}\mskip-1mu\rangle}
\def\spa#1.#2{\langle#1\,#2\rangle}
\def\spb#1.#2{[#1\,#2]}

\def\dk#1#2#3{
\begin{array}{r c l}
#1 & \rightarrow & #2 \\
 & & \bentarrow #3
\end{array}
}

\newcommand{\tc}{\tau\textsubscript{cut}}

\title{$H+1$~jet production revisited}

\author{John M. Campbell}
\email{johnmc@fnal.gov}
\affiliation{Fermilab, Batavia, IL 60510, USA}
\author{R. Keith Ellis}
\email{keith.ellis@durham.ac.uk}
\author{Satyajit Seth}
\email{satyajit.seth@durham.ac.uk}
\affiliation{Institute for Particle Physics Phenomenology,
Department of Physics, Durham University, Durham DH1 3LE, United Kingdom
\vspace*{1cm}}
\preprint{FERMILAB-PUB-19-189-T,\, IPPP/19/33}

\begin{abstract}
\vspace*{5pt}
We revisit the next-to-next-to-leading order (NNLO) calculation of 
the Higgs boson$+1$~jet production process, calculated in the $m_t \to \infty$
effective field theory. 
We perform a detailed comparison of the result calculated using the
jettiness slicing method, with published results obtained using
subtraction methods. The results of the jettiness calculation agree
with the two previous subtraction calculations at benchmark
points. The performance of the jettiness slicing approach is greatly
improved by adopting a definition of $1$-jettiness that accounts for the
boost of the Born system.  Nevertheless, the results demonstrate that
power corrections in the jettiness slicing method remain significant.
At large transverse momentum the effect of power corrections is much
reduced, as expected.
\end{abstract}
\maketitle

\section{Introduction}

Testing the properties of the Higgs boson is a central theme
of the experimental program of the LHC and will continue to be so
for the foreseeable future.  Despite the array of probes performed
so far, as yet no compelling evidence for unexpected couplings of the
Higgs boson to other particles has been discovered.  However, as more data
is accumulated, the experiments will be able to test our understanding
of the nature of the Higgs boson in interesting new ways.  One such
direction is through the production of a Higgs boson at non-zero transverse
momentum, a process mediated primarily by a Higgs boson recoiling against
one or more partons.  Such events contribute significantly to the total
number of Higgs boson events that can be observed.  This is due to the
copious radiation expected from the initial-state gluons that originate
the lowest-order inclusive production process.  Moreover, as the hardness of
the QCD radiation increases, partons are able to resolve the nature of the
loop-induced coupling and the process becomes sensitive to the particles that circulate
in the loop.  It is for this reason that measurements of Higgs boson
production in association with QCD radiation constitute a complementary
probe of the Higgs boson.

To turn such measurements into compelling information on the nature
of the Higgs boson requires precision theoretical calculations with which
to compare the experimental data.  At fixed order the description of such
events can be primarily described by the recoil of a Higgs boson against
a single jet, at least in a region of transverse momentum that is hard
enough to be properly described by a jet.  In order to achieve a suitable
precision, and a sufficiently small dependence on the unphysical
renormalization and factorization scales that enter the calculation, it
is necessary to perform computations up to next-to-next-to-leading order
(NNLO).  Over the last five years such predictions have become available
thanks to independent calculations from a number of groups~\cite{
Boughezal:2013uia,Chen:2014gva,Boughezal:2015dra,Boughezal:2015aha,
Caola:2015wna,Chen:2016zka}.  Beyond this, further steps have been
taken to also account for the effect of the resummation of
next-to-next-to-next-to-leading logarithms (N$^3$LL) to enable a better
description at small transverse momenta~\cite{Chen:2018pzu,Bizon:2018foh}.

The availability of multiple calculations of Higgs+jet production at NNLO
is important for a number of reasons.  First of all, the calculations have
been performed with a variety of different methods for handling
soft and collinear divergences in real radiation contributions.  The appearance
of such divergences leads to considerable complication in the calculations
and, depending on the details of the method, handling them could
expose the calculations to issues of numerical precision or systematic
flaws in the methods.  Second, to the extent that independent
calculations arrive at the same answer, additional confidence in the
theoretical calculations and methodologies is gained.  To understand these
issues it is important to benchmark the calculations appropriately and perform
detailed studies of any apparent disagreement.  For the case at hand,
a first comparison of results between the calculations was performed in the
context of studies for the LHC Higgs Cross Section Working Group
Yellow Report (``YR4'')~\cite{deFlorian:2016spz}.  A comprehensive
comparison was then performed by the NNLOJET group~\cite{Chen:2016zka} that
found agreement with the results of Refs.~\cite{Boughezal:2015dra,Caola:2015wna}
but was unable to confirm the results published in Ref.~\cite{Boughezal:2015aha}.
The latter result was obtained using the $N$-jettiness
method~\cite{Boughezal:2015dva,Gaunt:2015pea}, that relies on a factorization
theorem in Soft-Collinear Effective Theory (SCET) in order to compute a class
of unresolved contributions.  Therefore the resulting calculation closely resembles
a traditional slicing approach to higher-order corrections and is thus sensitive to the
value of a resolution parameter through the effect of power corrections to the
factorization formula.  To understand whether or not the difference could be
attributed to such effects, for instance as suggested in Ref.~\cite{Moult:2016fqy},
and to understand the effectiveness of the $N$-jettiness method more generally, requires
a detailed reappraisal of the calculation.  This paper aims to shed light on these
issues through our own implementation of the NNLO corrections to Higgs+jet production
using the $N$-jettiness method.

The outline of the paper is as follows.  In Section~\ref{sec:calc} we describe
the calculation and the various checks that have been performed on the ingredients.
A detailed comparison of results obtained using our calculation, and those of
NNLOJET~\cite{Chen:2014gva,Chen:2016zka,Bizon:2018foh}, follows in
Section~\ref{sec:nnlojet}.  We then compare results, under a different
set of cuts, with those of Ref.~\cite{Boughezal:2015dra} in Section~\ref{sec:bcmps}.
In Section~\ref{sec:boosted} we perform a study of the effectiveness of our
calculation in the boosted region and we conclude in Section~\ref{sec:conclusions}.


\section{Calculation}
\label{sec:calc}

Our $N$-jettiness calculation of Higgs+jet production is embedded in the
MCFM code~\cite{Campbell:2011bn,Campbell:2015qma}, with many ingredients in common
with previous NNLO calculations of color-singlet production~\cite{Boughezal:2016wmq}
and inclusive photon and photon+jet processes~\cite{Campbell:2016lzl,Campbell:2017dqk}.
In particular, all calculations share process-independent beam~\cite{Gaunt:2014xga,Gaunt:2014cfa}
and jet~\cite{Becher:2006qw,Becher:2010pd} functions.  We use the soft function calculation
of Ref.~\cite{Campbell:2017hsw}, which is in good agreement with two other evaluations
of the same quantity~\cite{Boughezal:2015eha,Bell:2018mkk}.  The remaining ingredient in
the SCET factorization theorem for the below-cut contribution is the hard function,
which we implement using the procedure of Ref.~\cite{Becher:2013vva} to obtain the
result up to 2-loop order using the helicity amplitudes of Ref.~\cite{Gehrmann:2011aa}.
The resulting hard function has been cross-checked against the result at a fixed kinematic
point that is also given in Ref.~\cite{Becher:2013vva}.

The remaining ingredient in the $N$-jettiness approach is the NLO calculation of the $H+2$~jet
process.  However, instead of applying the usual jet cuts, only a single jet is required and
additional parton configurations must pass a cut on $1$-jettiness.  This quantity is defined by,
\begin{equation}
{\mathcal T}_1 = \sum_m \min_i \left\{ \frac{2p_i \cdot q_m}{P_i} \right\} \,,
\label{eq:T1def}
\end{equation}
where the momenta $p_i$ are those of the partons in the initial beam and the (hardest) jet
that is present in the event, and the sum runs over the momenta of all partons, $q_m$.
A number of choices are possible for the normalization factors, $P_i$.  In this paper we will
always use the choice $P_i = 2 E_i$, resulting in a so-called geometric
measure~\cite{Jouttenus:2011wh,Jouttenus:2013hs}.  However, we will define ${\mathcal T}_1$ both
in the hadronic center-of-mass frame (as in previous $1$-jettiness calculations performed using
MCFM~\cite{Boughezal:2015ded,Campbell:2016jau,Campbell:2016yrh,Boughezal:2016wmq,Campbell:2016lzl,Campbell:2017dqk})
as well as in a boosted frame in which the system consisting of the Higgs boson
and the jet is at rest.  As explained in, for instance, Refs.~\cite{Gaunt:2015pea,Moult:2016fqy},
this is a more natural definition that should be less sensitive to power corrections at
large rapidities. 

Since the $H+2$~jet NLO calculation is used in a slightly different way than normal it should therefore
be scrutinized in detail.  In order to validate the helicity
amplitudes used in our calculation we have performed a cross-check of all matrix elements, 
contributing to both virtual and real contributions, against those obtained using
Madgraph5\_aMC@NLO~\cite{Alwall:2014hca} and found complete agreement.  To validate the proper
treatment of all singularities we have performed extensive checks of the subtraction terms in each
singular limit.  We have also limited the extent of all dipole subtractions using the introduction
of ``$\alpha$ parameters''~\cite{Nagy:1998bb} to test whether counter-terms have been included
consistently throughout the calculation.  

Up to now, $\alpha$-independence had typically only been checked for the total cross-section, usually
varying all parameters at the same time.  This hides potential deviations in sub-leading channels and can mask
mismatches in color orderings since, for example, in some channels color orderings do not matter
once final-initial and final-final dipoles are summed over. In order to provide a more stringent check
on the calculation we computed the $\alpha$-dependence for each partonic initial state and also for
each possible $\alpha$ parameter individually.  These checks revealed a small inconsistency in the subtraction
of singularities in the $q\bar q \to Hggg$ channel, and an even smaller discrepancy in
$qg \to H q q \bar q$ (identical-quark) contributions.  Together, these effects resulted in $\alpha$-dependence
at a very small level in the total cross-section that had not been detected previously.  

\begin{figure}
\begin{center}
\includegraphics[width=0.45\textwidth]{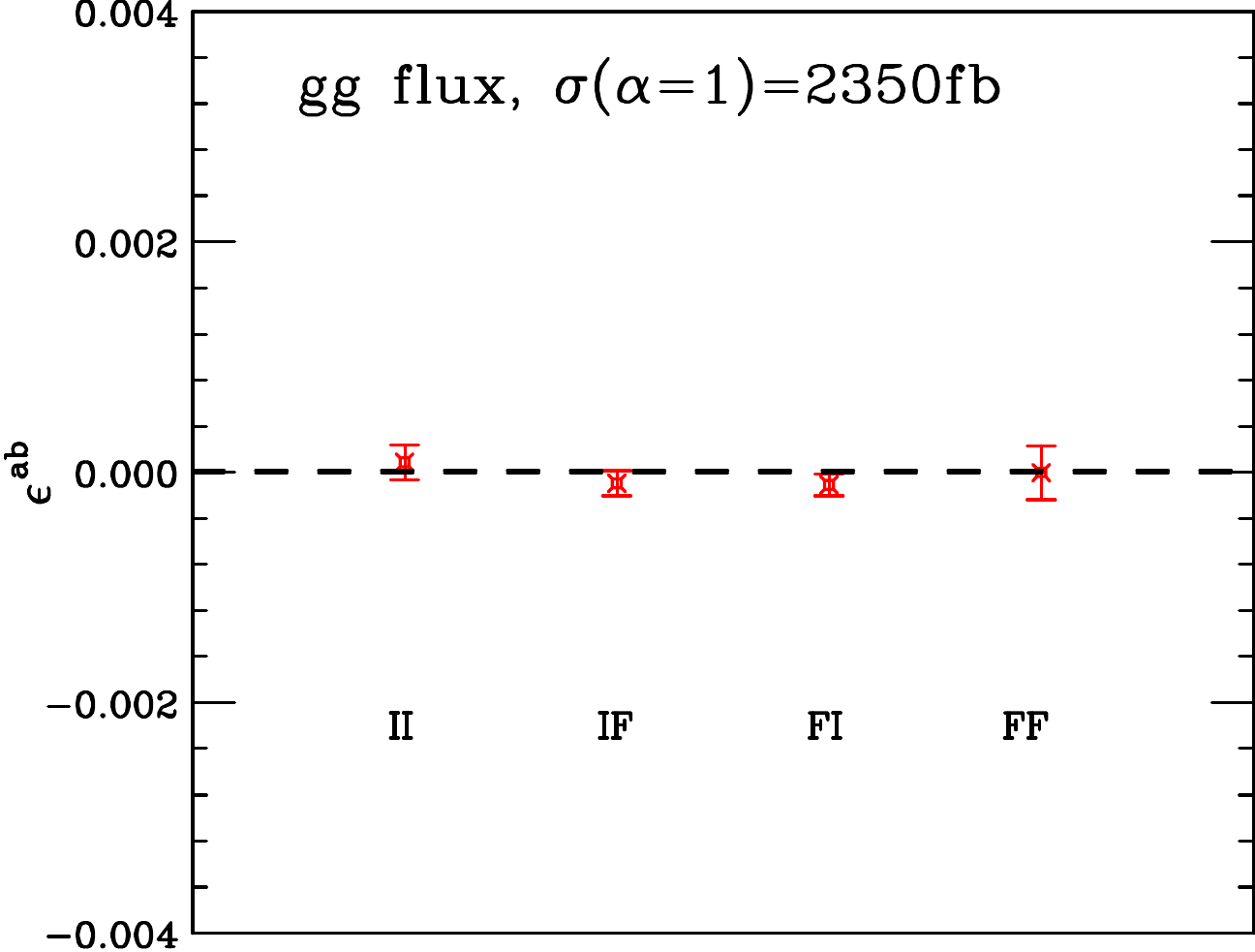} ~
\includegraphics[width=0.45\textwidth]{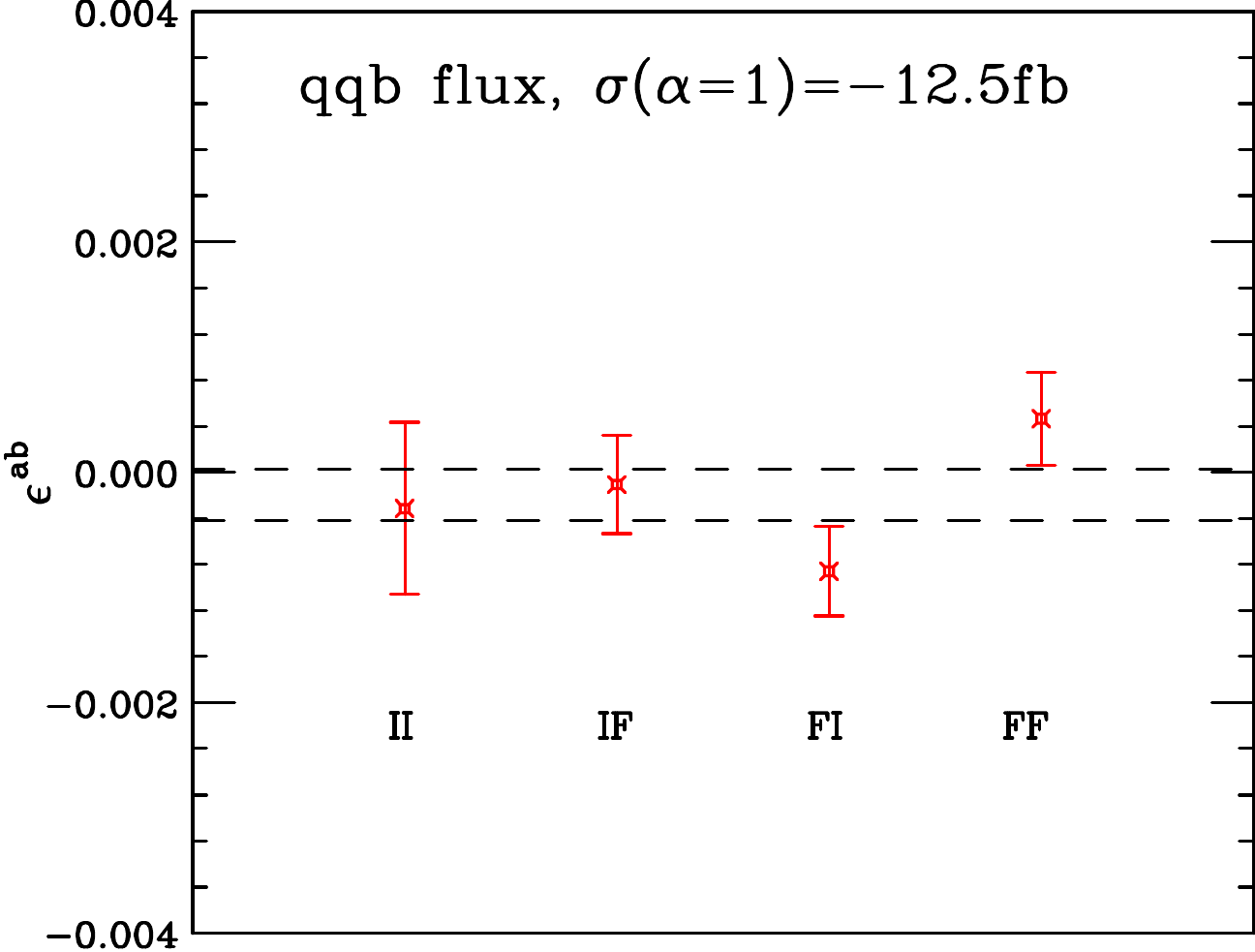} \\ \vskip2mm
\includegraphics[width=0.45\textwidth]{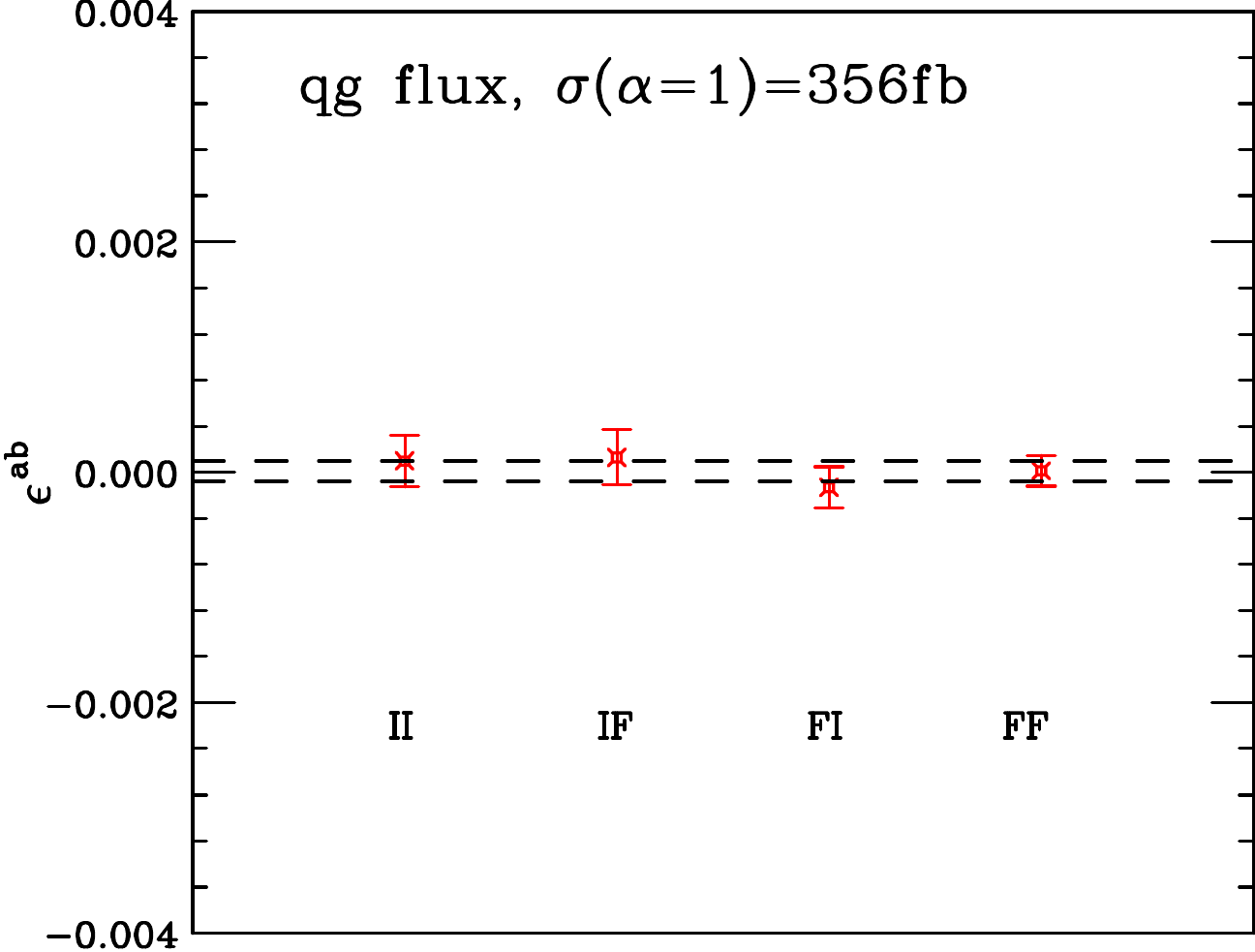} ~
\includegraphics[width=0.45\textwidth]{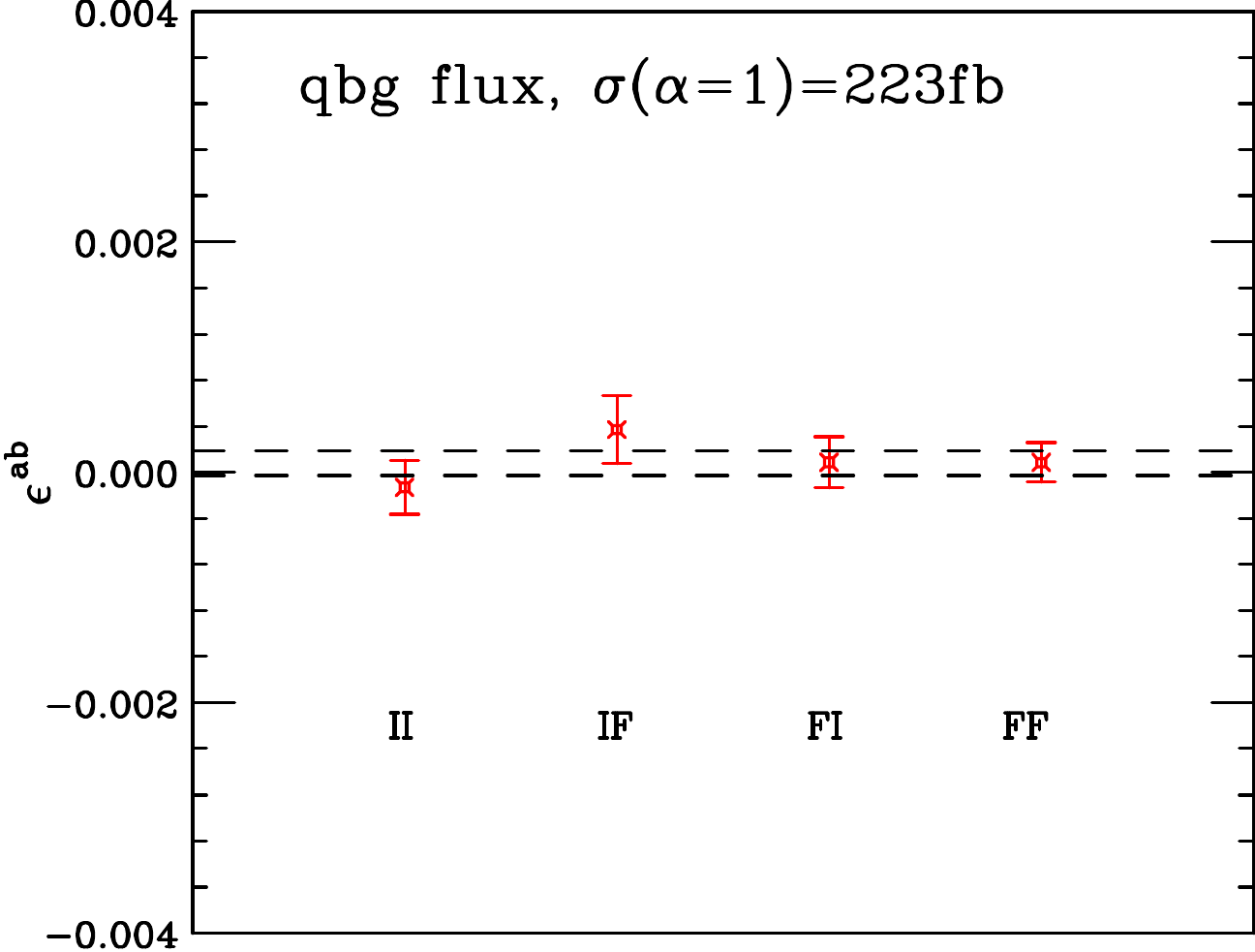} \\ \vskip2mm
\includegraphics[width=0.45\textwidth]{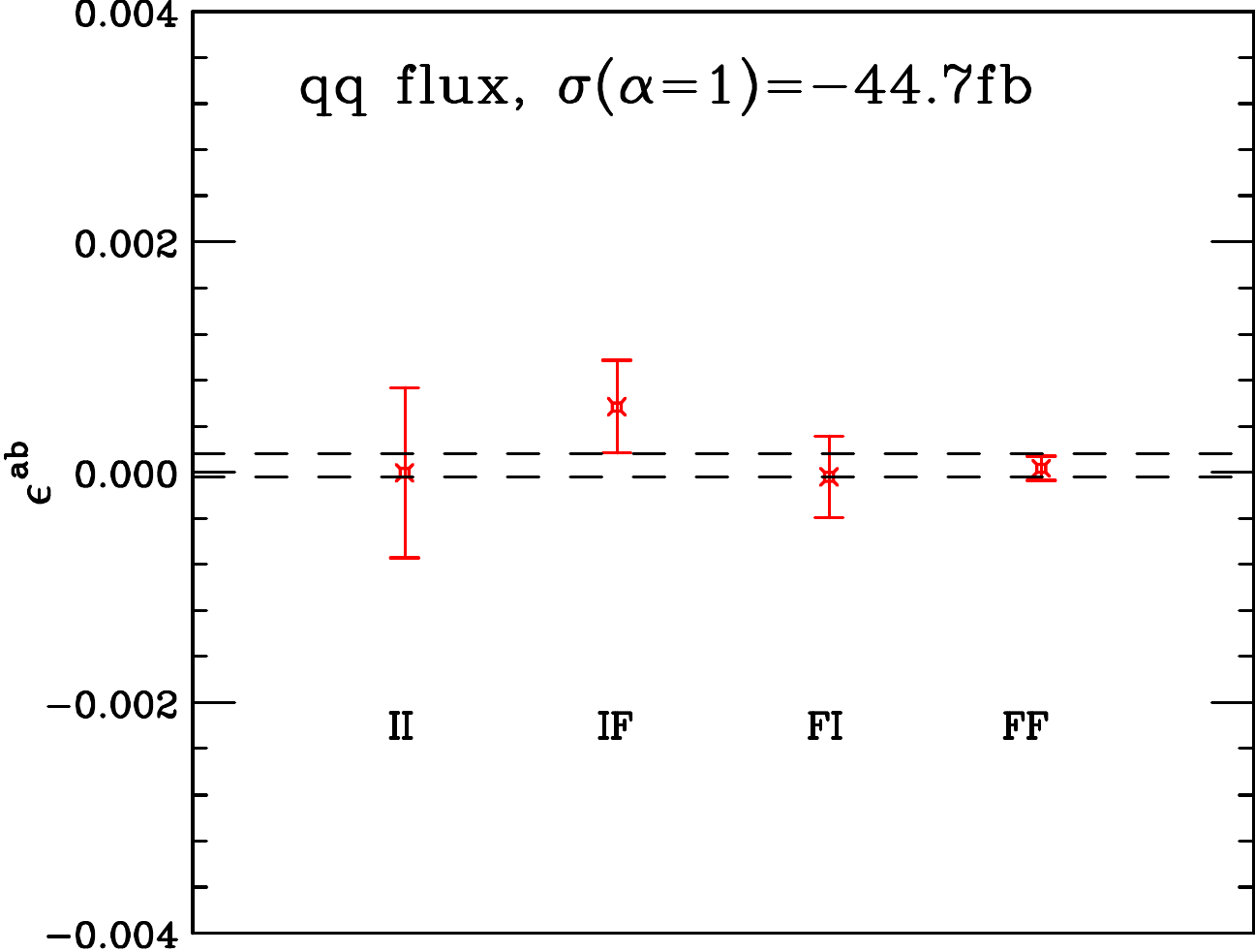} ~
\includegraphics[width=0.45\textwidth]{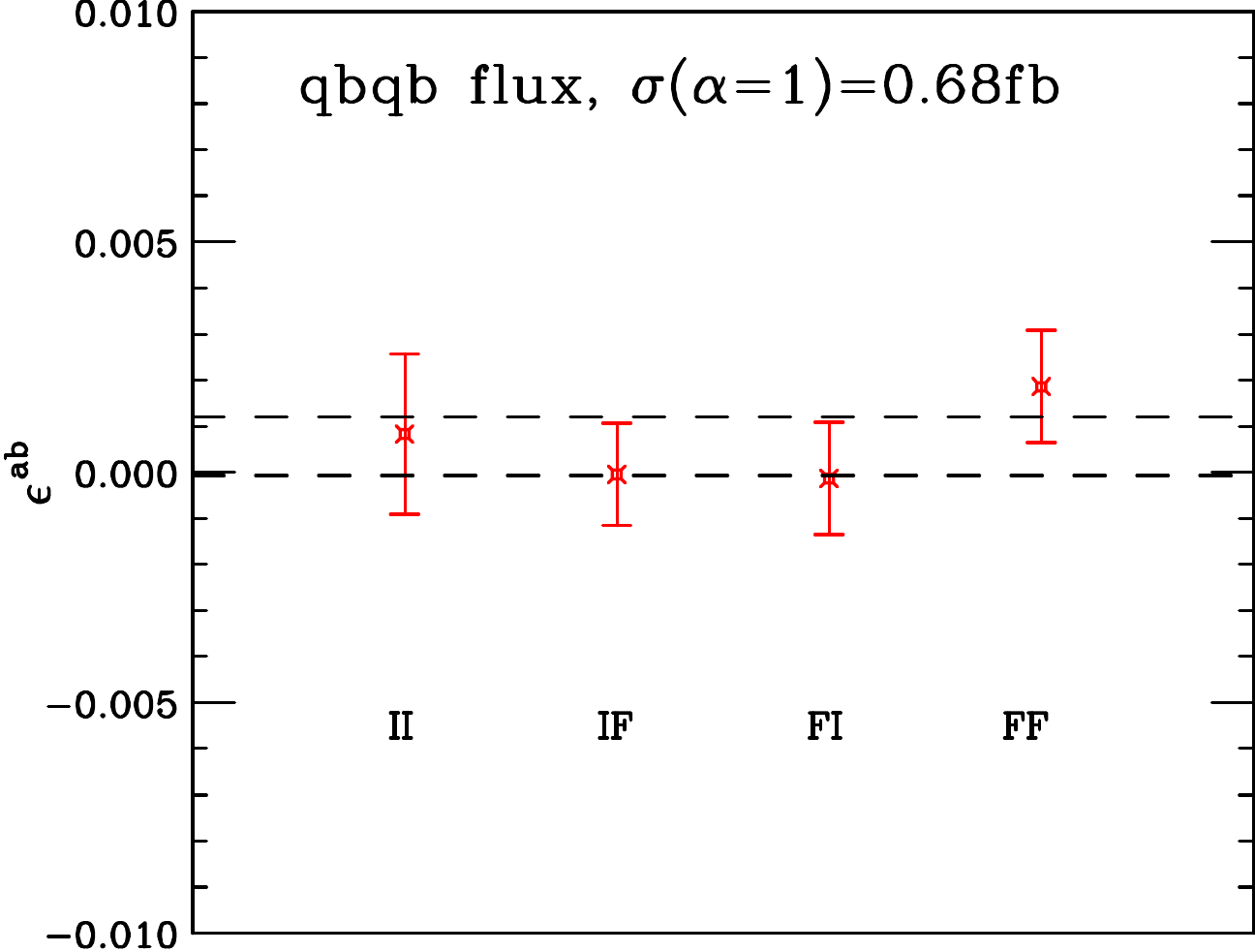}
\end{center}
\caption{The dependence of the $H+2j$ cross-section on the $\alpha$ parameters, for each of the
different partonic fluxes.
The points represent the deviation from the default ($\alpha_{II}=\alpha_{IF}=\alpha_{FI}=\alpha_{FF}=1$)
when the labelled parameter is set to $10^{-2}$.  The cross-sections in this channel, obtained using
the default parameters, are indicated in the plots.  The dashed lines represent the uncertainty on
a fit of the results to a constant, indicating excellent agreement with zero at the level of Monte
Carlo statistics.}
\label{fig:alphadep-all}
\end{figure}

To illustrate the level of $\alpha$-independence in the code used for the present calculation we
will show the results of cross-checks performed using the following setup:
\begin{eqnarray}
{\rm LHC},~\sqrt{s} = 13~{\rm TeV}, \qquad && \mu_R = \mu_F = m_H = 125~{\rm GeV}, \nonumber \\
p_T^{\rm jet} > 20~{\rm GeV}, \qquad && \Delta R = 0.4
\end{eqnarray}
Jets are clustered according to the anti-$k_T$ algorithm and, as indicated above, no explicit
cut on their rapidities is applied.
The results are shown in Figure~\ref{fig:alphadep-all}, which indicates the deviation from the default
($\alpha_{II}=\alpha_{IF}=\alpha_{FI}=\alpha_{FF}=1$) when each of the dipole parameters is set to
$10^{-2}$.  The deviation is measured by,
\begin{equation}
\epsilon^{ab} = \frac{\sigma(\alpha_{ab} = 1) - \sigma(\alpha_{ab} = 0.01)}{\sigma(\alpha_{ab} = 1)} \,.
\end{equation}
Note that, when going between these two values of $\alpha$, the virtual and real contributions each
individually change by an amount that often far exceeds the total cross-section itself so that
the check is a rather stringent one.
The results in Figure~\ref{fig:alphadep-all} show that the cross-section is
independent of the choice of $\alpha$ parameters, over a wide range, to within the Monte Carlo statistics
indicated for each channel.  This corresponds to a check at the $0.1\%$ level or better for all channels
except $\bar q \bar q$, where the size of the cross-section is so small that the check is slightly less
strict, at the $0.3\%$ level.  Since, in general, the calculation is more efficient for $\alpha < 1$ we
choose to set $\alpha_{II}=\alpha_{IF}=\alpha_{FI}=\alpha_{FF}=0.01$ to obtain all the results presented
hereafter.

Beyond the issues discussed above, the use of the $H+2$~jet NLO process in a $1$-jettiness calculation
requires a number of small further refinements.  First, the evaluation of the real corrections probes
partonic configurations that can become highly singular, particularly for very small values
of the $1$-jettiness cut.  This means that special attention must be paid to generating
phase-space points in this region.  Moreover, at NLO it is typical to implement a technical cut
in order to remove extreme phase-space configurations in which the real emission matrix element and 
subtraction counter-terms should exactly cancel, but for which numerical
stability can be an issue.  In the NNLO calculation it is important to ensure that any such
cut does not impact the result, which typically requires the cuts to be made at smaller values than
in a typical NLO calculation.  We have performed detailed checks to ensure that, with the technical
cuts that we have used, points that are removed do not alter our results.  Finally, the NLO code must
be modified trivially in order to properly account for all higher-order corrections to the Wilson
coefficient that couples the Higgs field to two gluons in the effective field theory~\cite{Dawson:1990zj,Chetyrkin:1997iv}.


\section{Comparison with NNLOJET}
\label{sec:nnlojet}

We now turn to a detailed comparison with the NNLO results provided by
NNLOJET~\cite{Chen:2014gva,Chen:2016zka,Bizon:2018foh},
employing the setup that was used for the YR4 comparison~\cite{deFlorian:2016spz}.\footnote{
We thank Xuan Chen and Nigel Glover for instigating this comparison and for providing
a detailed breakdown of their results that is used here.}
These are summarized here:
\begin{eqnarray}
{\rm LHC},~\sqrt{s} = 13~{\rm TeV}, && \quad \mu_R = \mu_F = m_H = 125~{\rm GeV}, \nonumber \\
p_T^{\rm jet} > 30~{\rm GeV}, && \quad {\rm anti-}k_T~{\rm algorithm},~\Delta R = 0.4 \\
{\rm PDF~set:} && \quad {\tt PDF4LHC15\_nnlo\_30} \nonumber
\end{eqnarray}
Note that this choice of PDF set is used to obtain results both at NLO and NNLO.
By inspecting these cuts one might already anticipate a potential disadvantage
to using the jettiness slicing method for the calculation of NNLO corrections.  This
is because neither the jet nor the Higgs boson is required to satisfy any rapidity
constraint, leading to contributions to the cross-section from events with high-rapidity
particles.  These types of event have already been identified as being subject to power corrections
that are large~\cite{Ebert:2018lzn}.

Up to NNLO in QCD, the cross-section for this process can be written as,
\begin{equation}
\sigma_{NNLO} = \sigma_{LO} + \delta\sigma_{NLO} + \delta\sigma_{NNLO},
\label{eq:orders}
\end{equation}
where $\sigma_{LO}$, $\delta\sigma_{NLO}$ and $\delta\sigma_{NNLO}$ contain, respectively,
only contributions of order $\alpha_s^3$, $\alpha_s^4$ and $\alpha_s^5$.  The NLO
cross-section, $\sigma_{NLO}$, is defined similarly by omitting the final term.
In the sections that follow it is useful to compare calculations of both the higher-order
coefficients $\delta\sigma_{NLO}$ and $\delta\sigma_{NNLO}$ as well as the full cross-sections
at each order, $\sigma_{NLO}$ and $\sigma_{NNLO}$.

\subsection{Comparison of NLO calculation}
We have first cross-checked the implementation of the NLO calculation, using dipole subtraction, by comparing with the corresponding
computation in NNLOJET.  As shown in Table~\ref{tab:h1j-nlo}, we have found complete agreement between the codes at the per-mille level.

\begin{table}[!h]
\begin{tabular}{@{}lllllll@{}}
\hline
NLO calculation    ~  & ~~~$gg$~~          & ~~~$qg$~~           & ~~~$\bar{q}g$~~     & ~~~$q\bar{q}$~~       & ~~~$qq$~~             & ~~~$\bar{q}\bar{q}$~~ \\ 
\hline
NNLOJET  & $ 4962 \pm  3 $~~ & $  546.6 \pm 0.6$~~& $ 231.5 \pm 0.2$~~ & $ -14.61 \pm 0.03$~~ & $ -34.01 \pm 0.13$~~ & $ -6.739 \pm 0.008 $ \\
MCFM     & $ 4960 \pm  2 $ & $  546.3 \pm 0.4$& $ 231.1 \pm 0.1$ & $ -14.62 \pm 0.04$ & $ -33.94 \pm 0.08$ & $ -6.731 \pm 0.011 $ \\
\hline
\end{tabular}
\caption{The NLO contribution $\delta_{NLO}$, defined in Eq.~\ref{eq:orders}, broken down into individual partonic
channels, as computed by NNLOJET and MCFM (dipole subtraction).  Cross-sections are shown in femtobarns.}
\label{tab:h1j-nlo}
\end{table}

We now turn to the $1$-jettiness calculation and inspect the $\tc$ dependence of each partonic channel, using a value of $\tc$ that
depends dynamically on the kinematics of each event.  Specifically,
we set
\begin{equation}
\tc = \epsilon \times \sqrt{ m_H^2 + \left(p_T^{j_1}\right)^2}
\label{eq;dynamictau}
\end{equation}
with $2 \times 10^{-5} \leq \epsilon \leq 5 \times 10^{-4}$.
For the sake of comparison it is possible to convert these values of
$\tc$ to a definite scale by using $p_T^{j_1} \to p_{T,\text{min}}^{j_1}$. In this way,
these values of $\epsilon$ approximately correspond to fixed values of $\tc$ in the range
$0.0025$~--~$0.06$~GeV, although the correspondence is not exact due to 
contributions to the cross-section at higher jet transverse momentum.
We note in passing that almost the entire range of $\tc$ studied here is significantly
below the one studied in the previous calculation of $H+$jet production using
jettiness slicing~\cite{Boughezal:2015aha}.

\begin{figure}
\begin{center}
\includegraphics[width=0.45\textwidth]{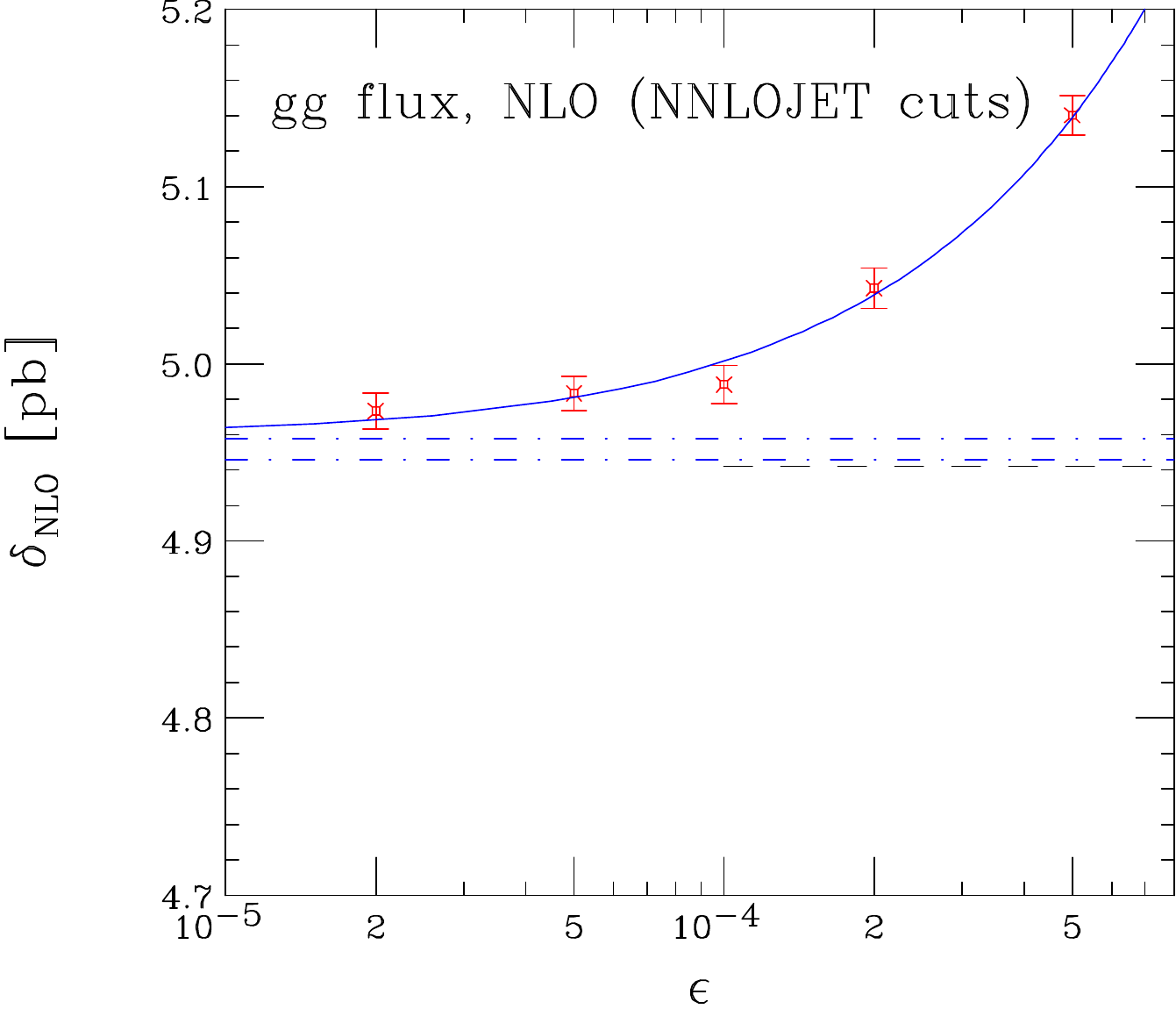}\hspace*{0.5cm}
\includegraphics[width=0.45\textwidth]{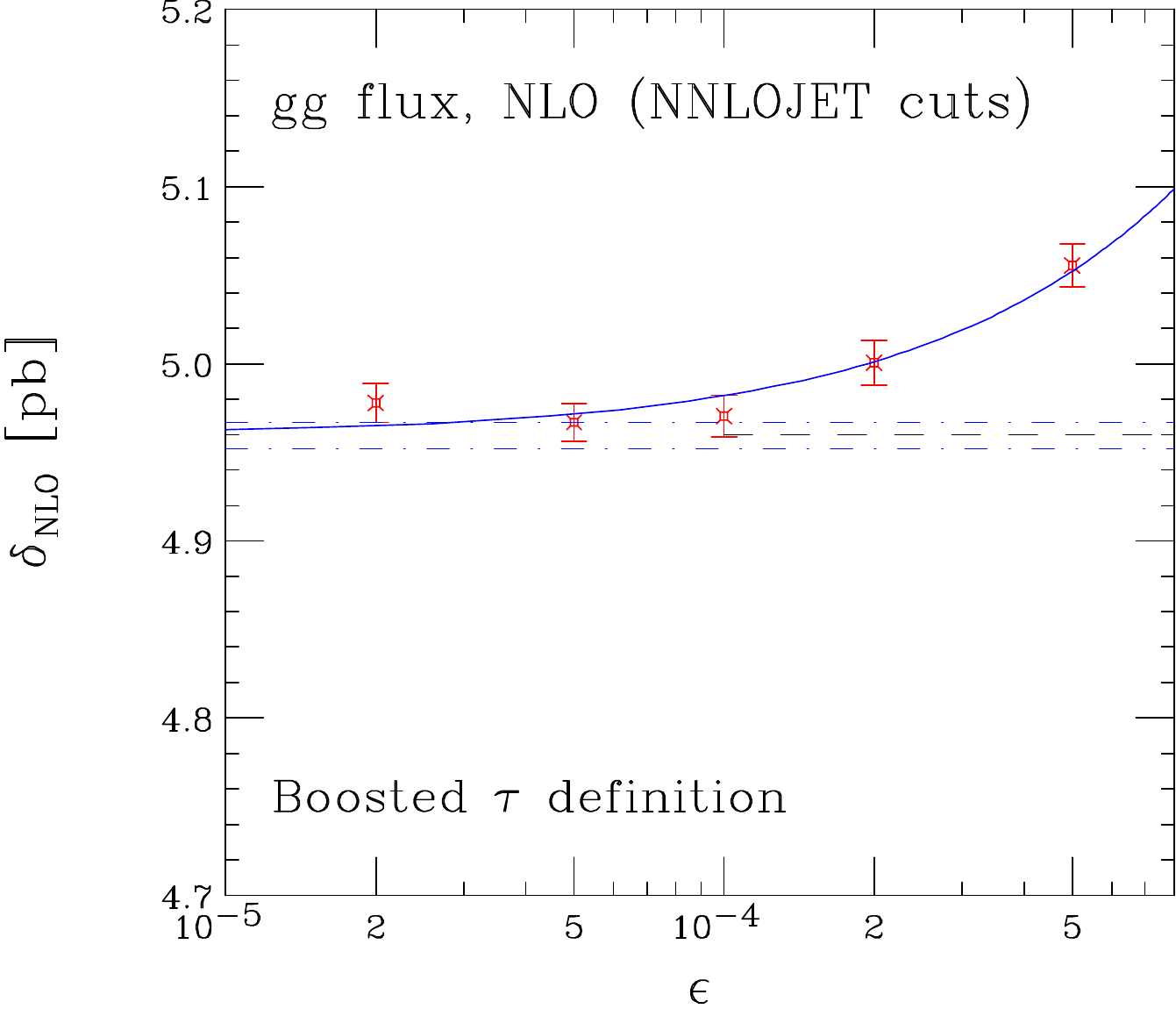} \\ \vspace*{0.45cm}
\includegraphics[width=0.45\textwidth]{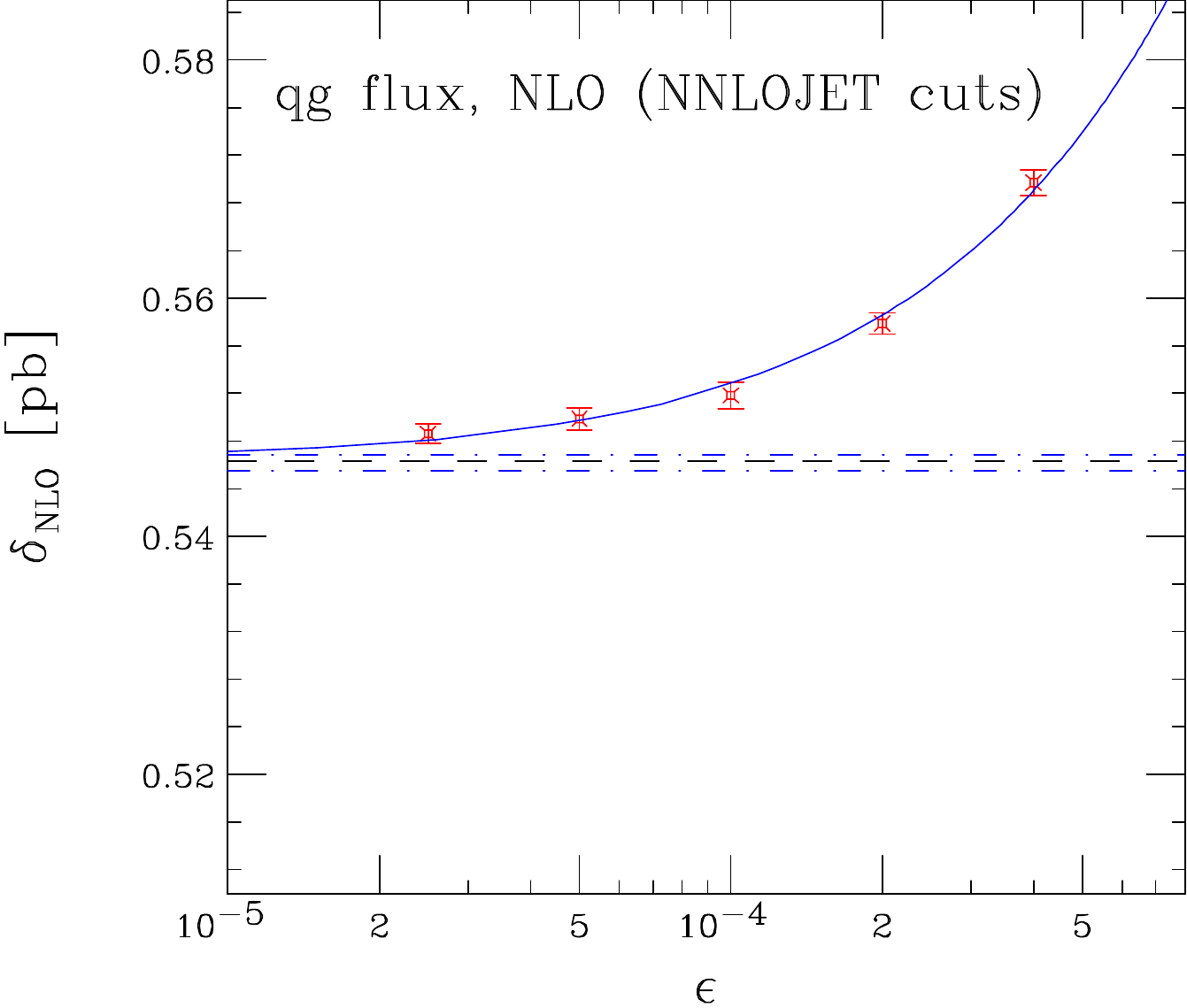}\hspace*{0.5cm}
\includegraphics[width=0.45\textwidth]{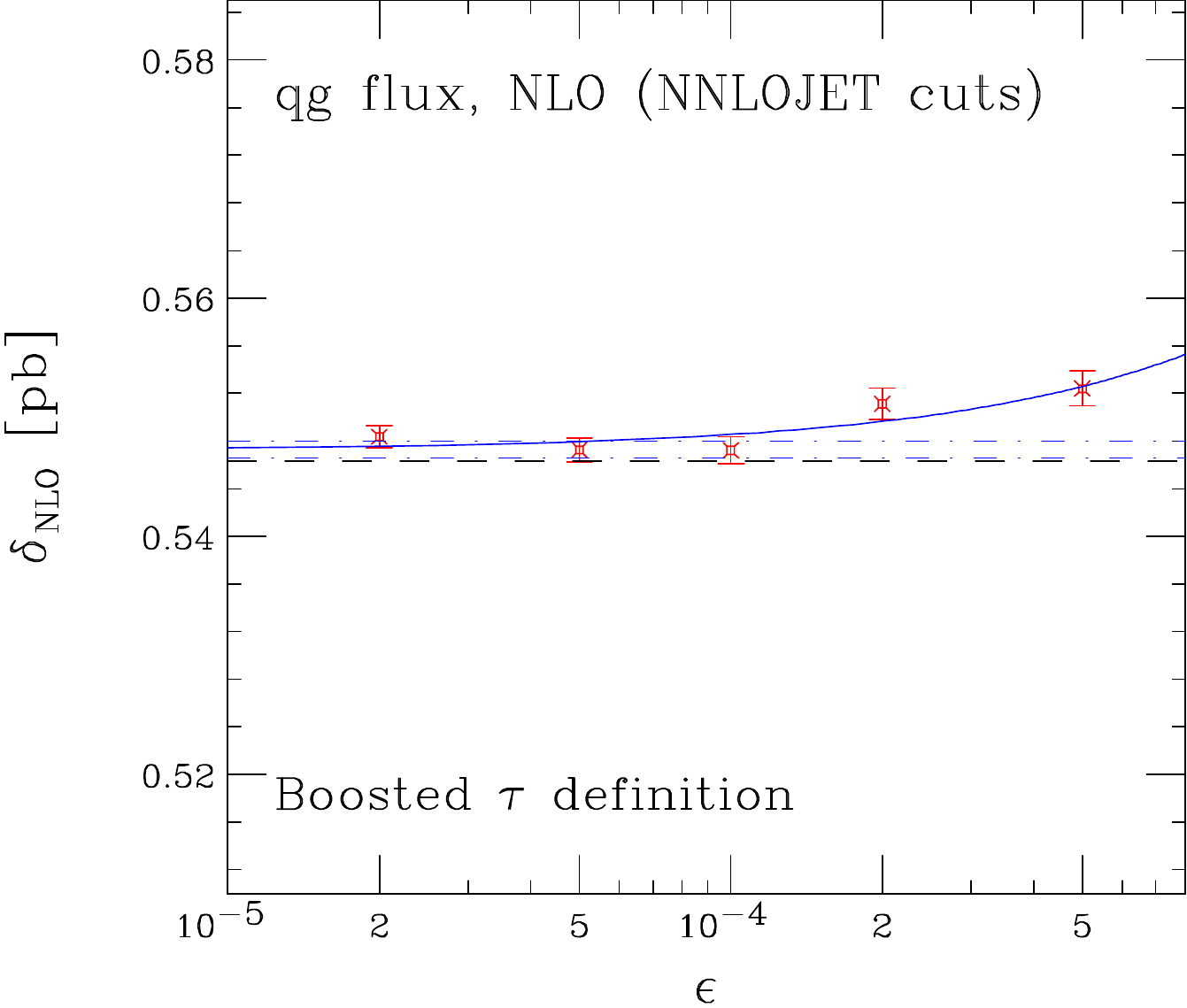} \\ \vspace*{0.45cm}
\includegraphics[width=0.45\textwidth]{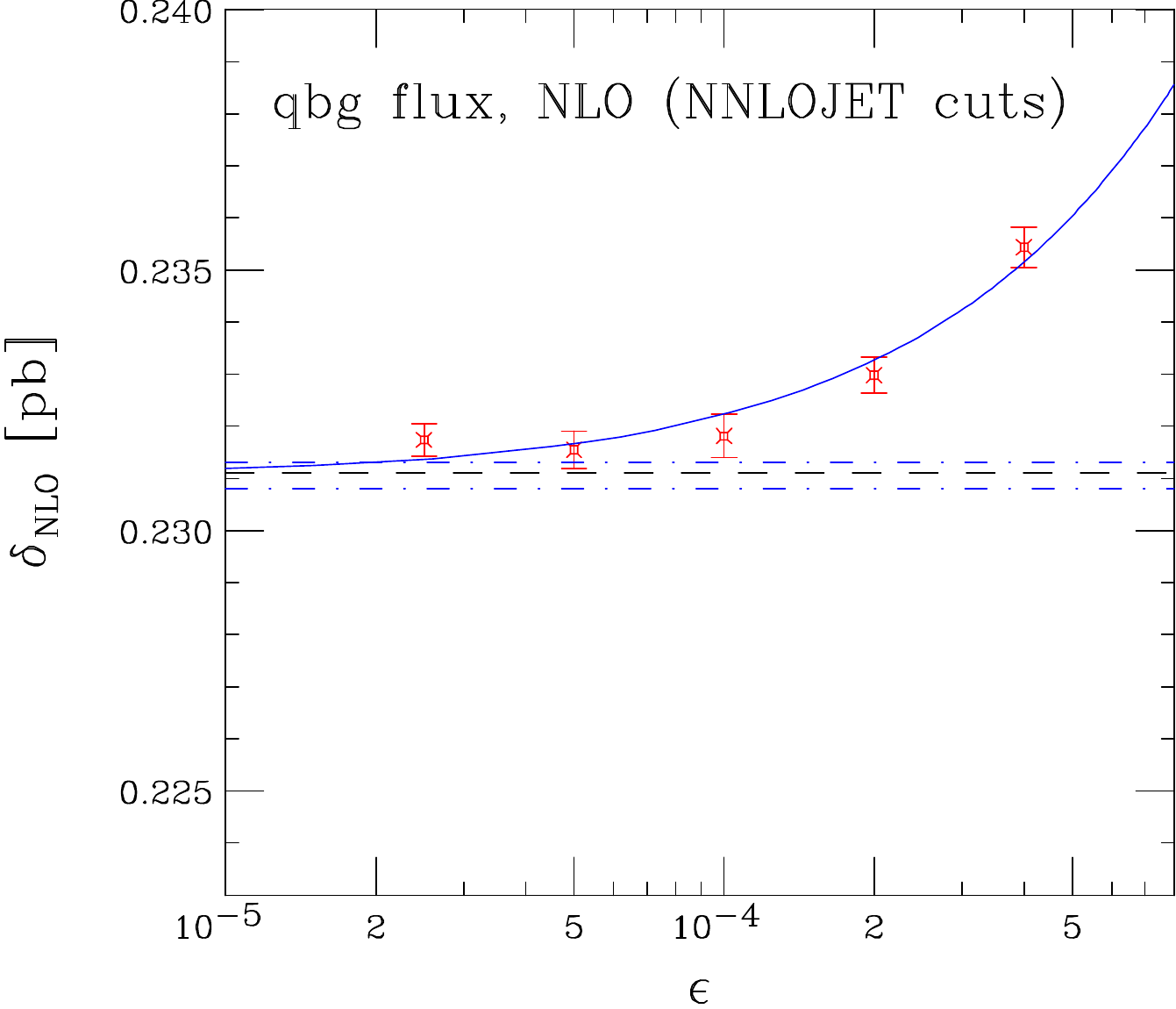}\hspace*{0.5cm}
\includegraphics[width=0.45\textwidth]{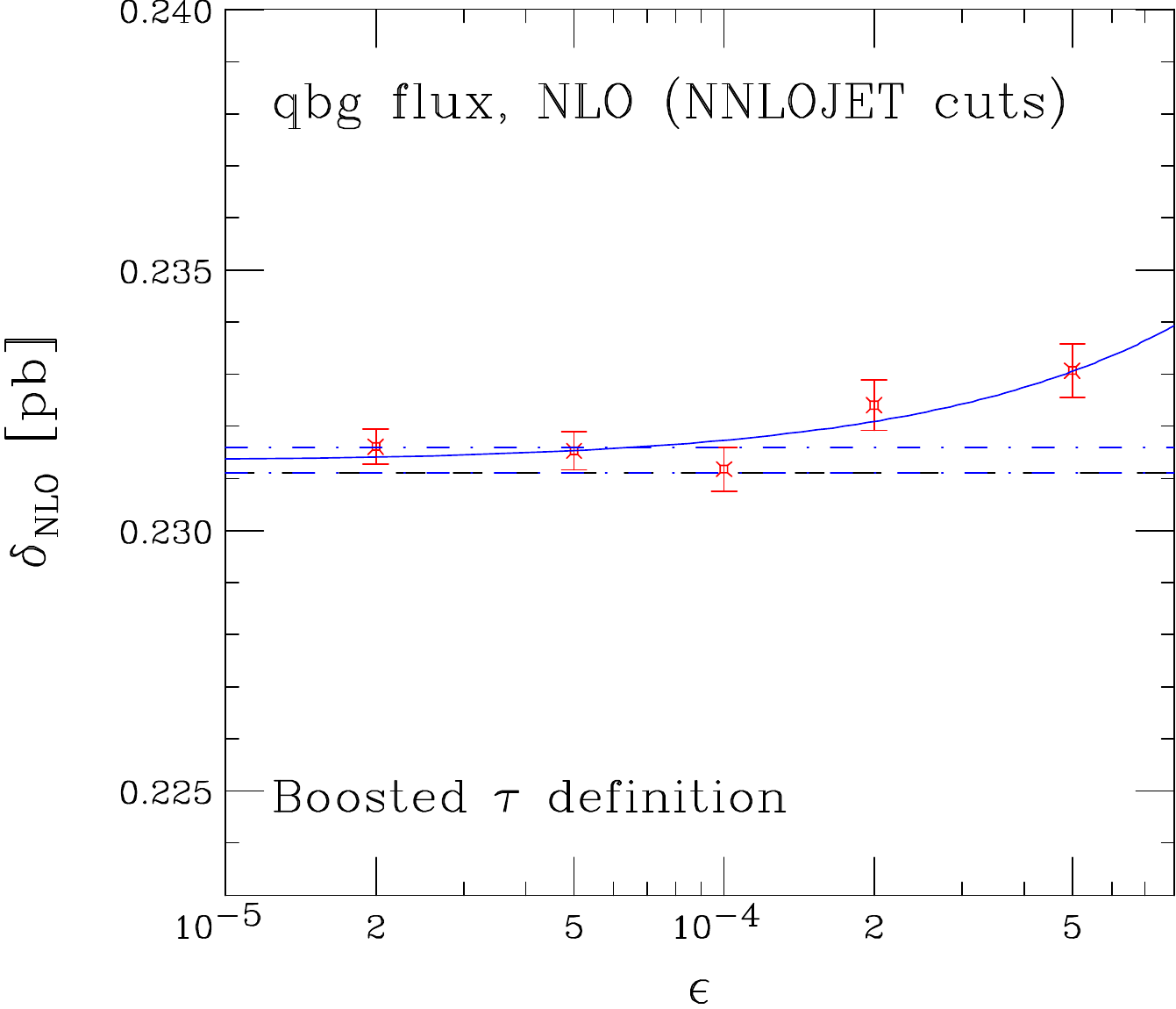}
\end{center}
\caption{$\tau$-dependence of NLO coefficients for the $gg$, $qg$ and $\bar q g$ partonic channels, in the NNLOJET setup.
The plots on the left show the result when ${\mathcal T}_1$ is computed in the hadronic c.o.m. and the ones on the right
indicate the corresponding result when evaluating this quantity in the boosted frame.
The (blue) solid lines correspond to the fit form in Eq.~(\ref{eq:fitform-nlo1}), with the dot-dashed lines representing
the errors on the asymptotic value of the fit.  The exact results, computed in MCFM using dipole subtraction,
are shown as the black dashed lines.}
\label{fig:taudep-nnlojet-nlo-leading}
\end{figure}

\begin{figure}
\begin{center}
\includegraphics[width=0.45\textwidth]{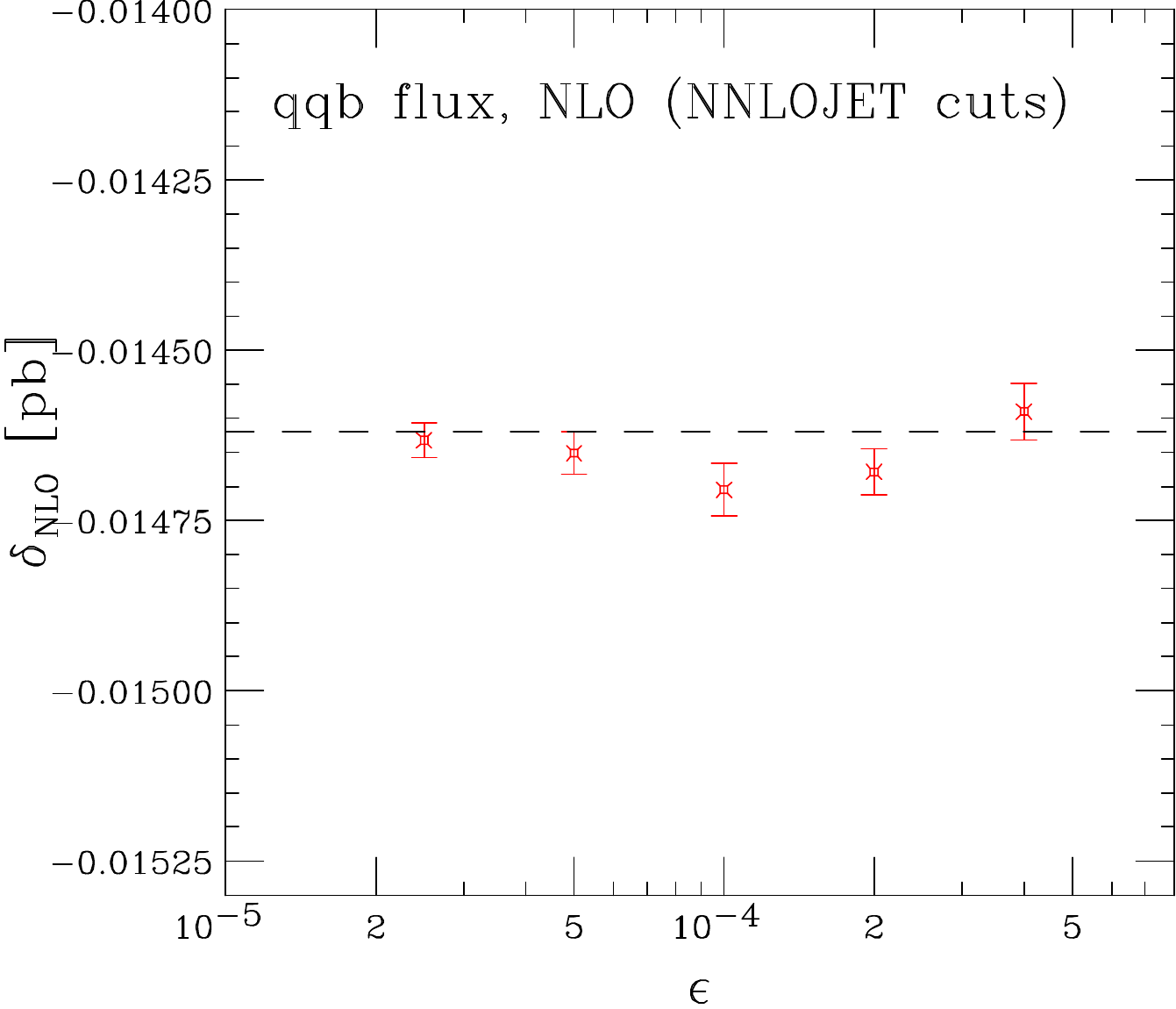}\hspace*{0.5cm}
\includegraphics[width=0.45\textwidth]{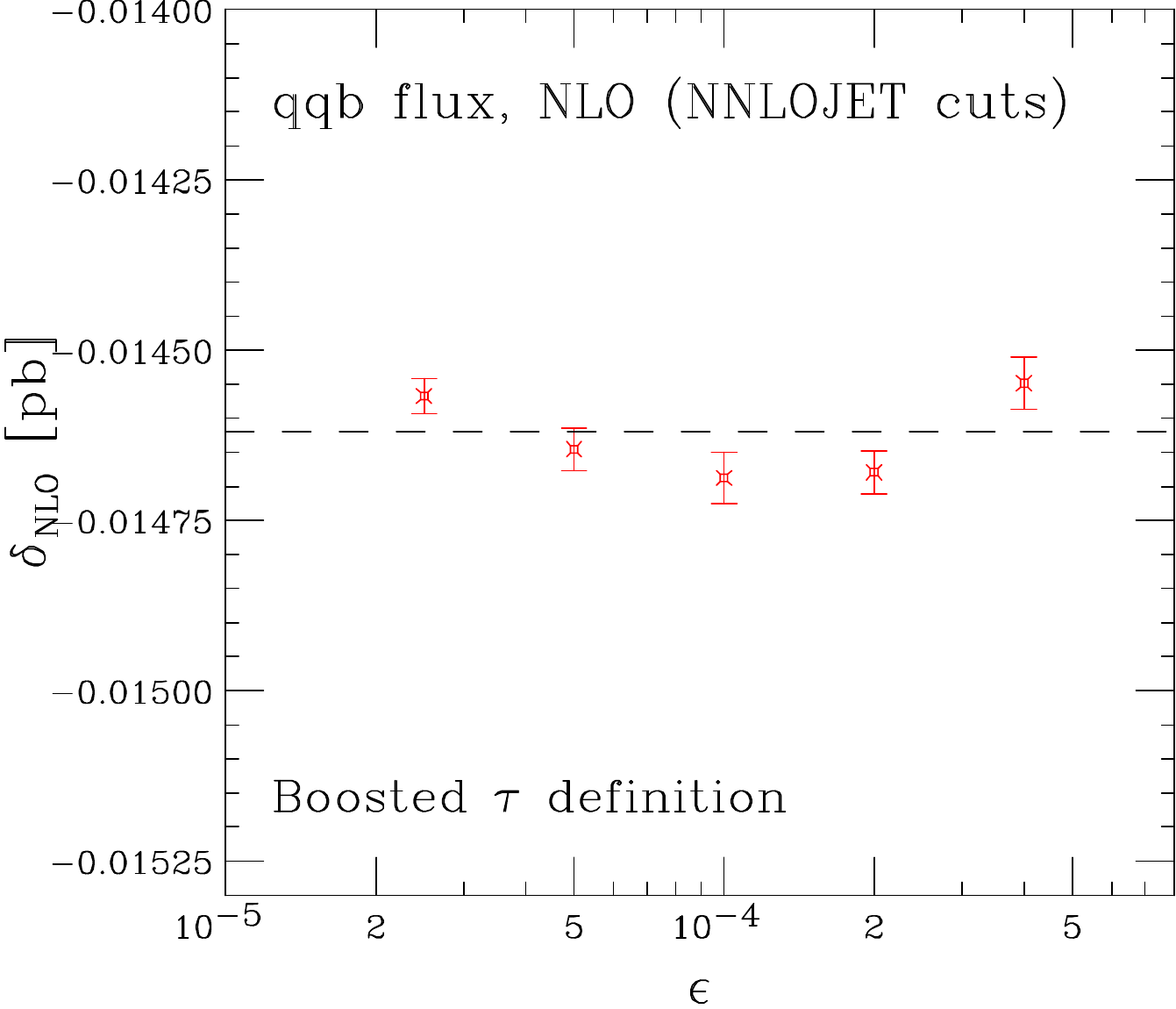} \\ \vspace*{0.5cm}
\includegraphics[width=0.45\textwidth]{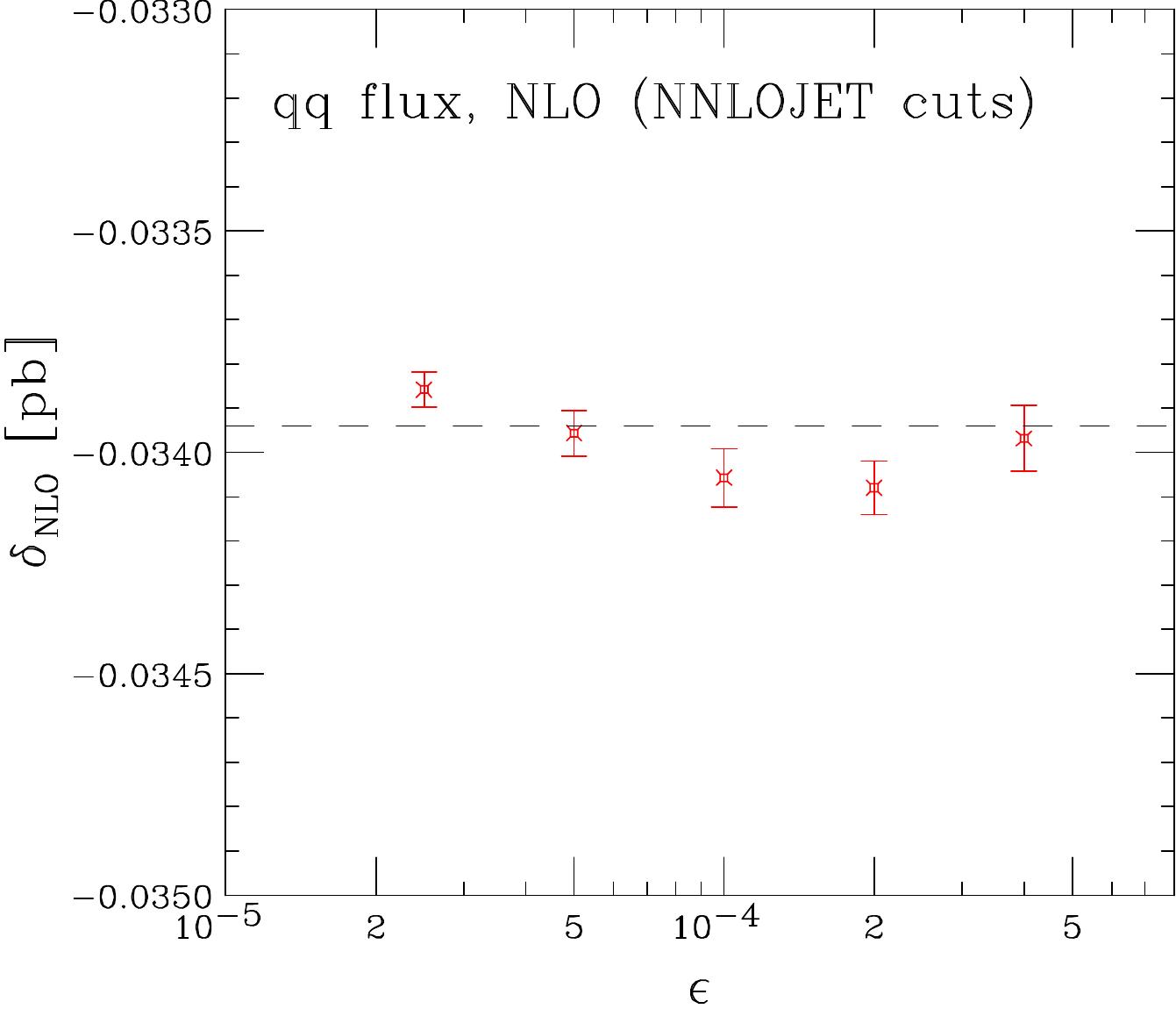}\hspace*{0.5cm}
\includegraphics[width=0.45\textwidth]{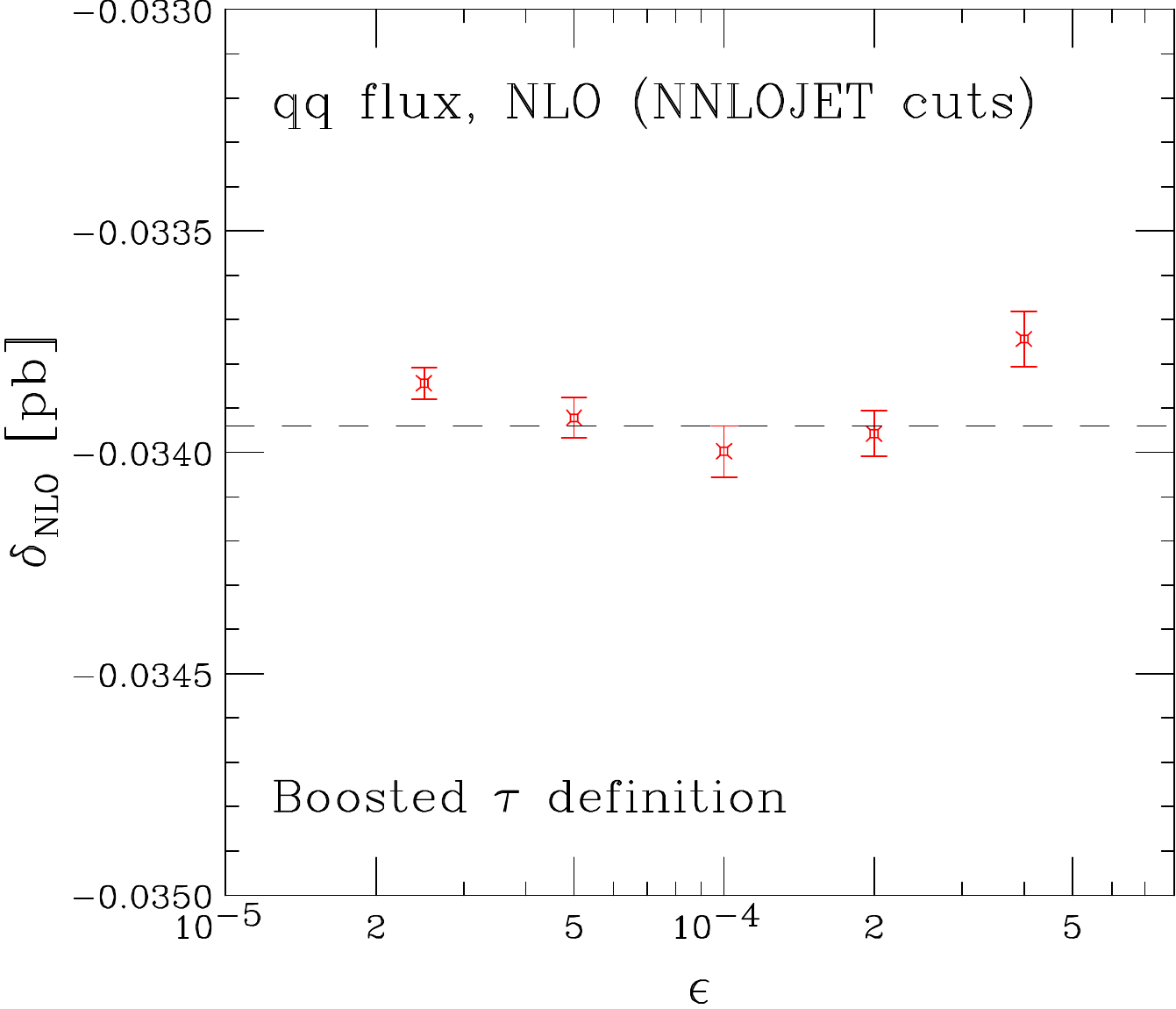} \\ \vspace*{0.5cm}
\includegraphics[width=0.45\textwidth]{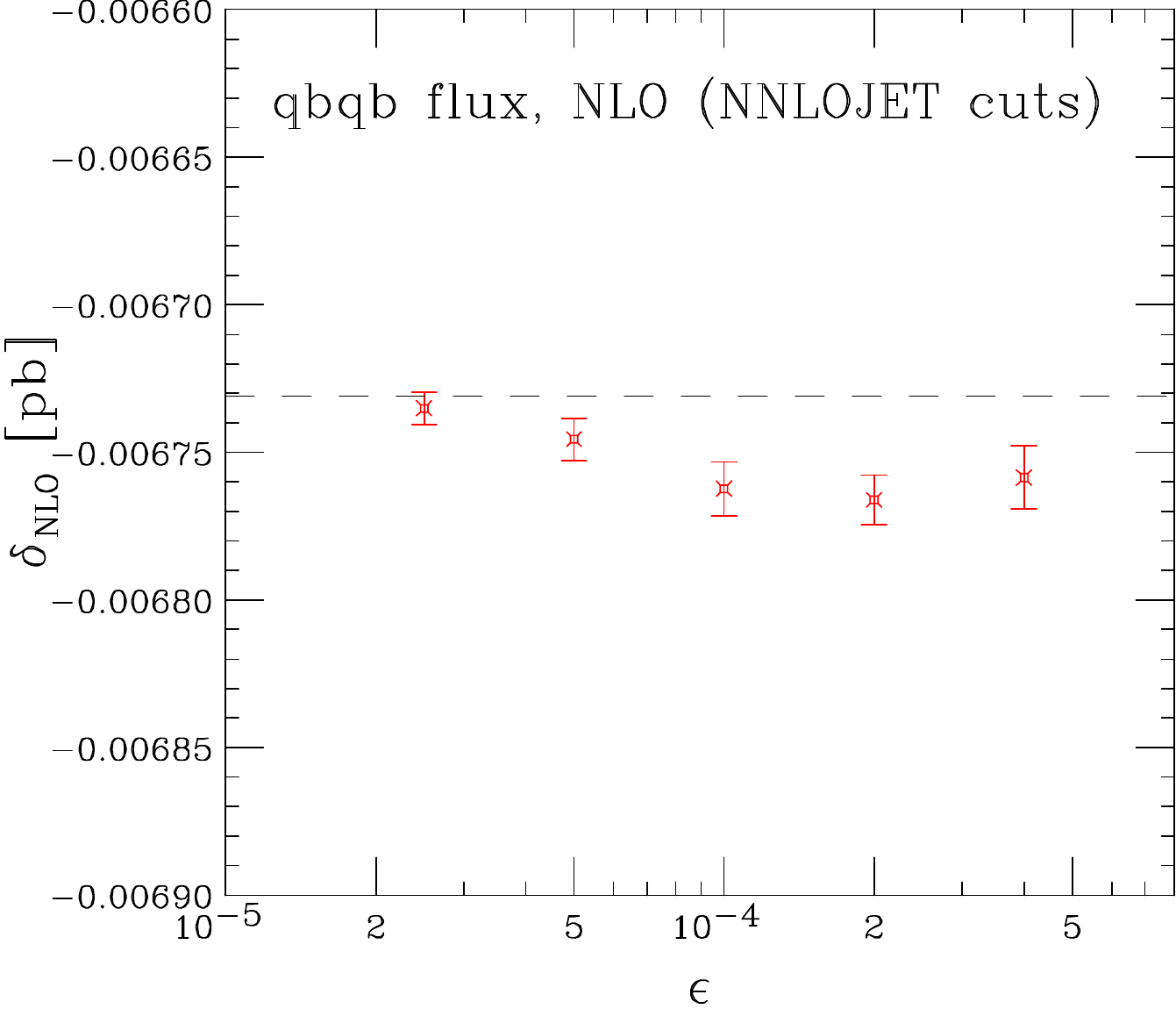}\hspace*{0.5cm}
\includegraphics[width=0.45\textwidth]{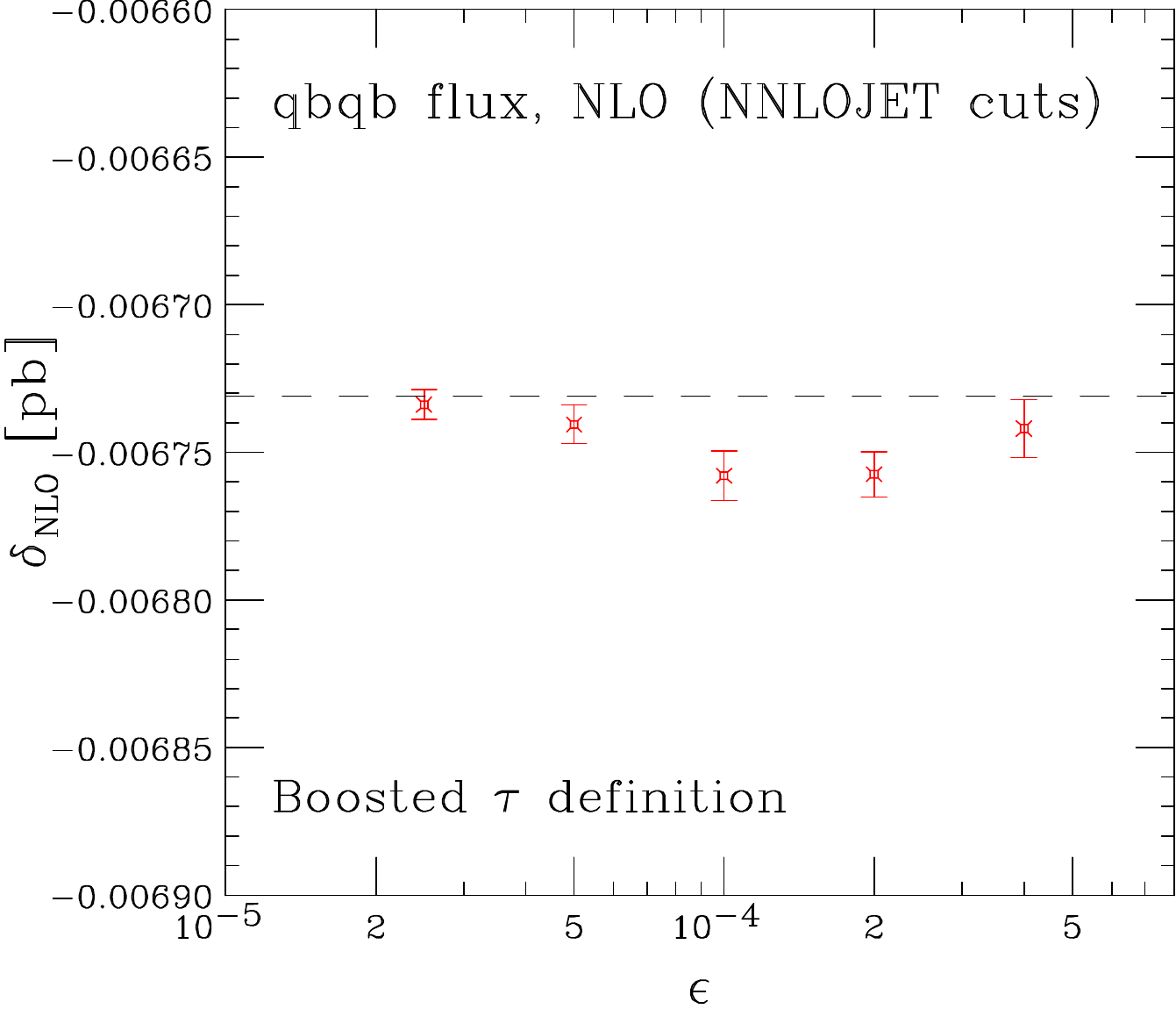}
\end{center}
\caption{$\tau$-dependence of NLO coefficients for the $q \bar q$, $qq$ and $\bar q \bar q$ partonic channels, in the NNLOJET setup.
The plots on the left show the result when ${\mathcal T}_1$ is computed in the hadronic c.o.m. and the ones on the right
indicate the corresponding result when evaluating this quantity in the boosted frame.
The exact results, computed in MCFM using dipole subtraction, are shown as the black dashed lines.
}
\label{fig:taudep-nnlojet-nlo-subleading}
\end{figure}

As a first check of the sensitivity of this process to power corrections, we examine
the  $\tc$ dependence of the NLO calculation in each of the three main partonic channels --
$gg$, $qg$ and $\bar q g$.  The results are shown in Fig.~\ref{fig:taudep-nnlojet-nlo-leading}, for
both definitions of ${\mathcal T}_1$, in the hadronic center-of-mass frame (left) and after
the boost to the rest frame of the Higgs boson+jet system (right). We see that, in both cases, the jettiness result for the NLO
coefficient in each channel approaches the known NLO result computed using dipole subtraction
as $\tc \to 0$.  However we also observe that, as expected, this approach is much less steep when using
the boosted definition of ${\mathcal T}_1$.   In order to quantify the $\tc$-dependence we have performed a fit to the data points 
using the expected behavior of the power corrections.  This is prescribed by the leading singularities at this order and
takes the form,
\beq
\delta_{NLO}^{\{gg, qg, \bar qg\}}(\epsilon) = \delta_{NLO}^{fit} + c_0 \, \epsilon \log(\epsilon) + \ldots
\label{eq:fitform-nlo1}
\eeq
These fits, shown as solid lines in Fig.~\ref{fig:taudep-nnlojet-nlo-leading}, describe the $\tc$-dependence extremely
well.  Corresponding results for the subleading channels -- $q \bar q$, $qq$ and $\bar q \bar q$ -- are shown in
Fig.~\ref{fig:taudep-nnlojet-nlo-subleading}.   Again we observe excellent agreement with the exact calculation.
However, from this figure it is obvious that the power corrections in these channels
are tiny, with agreement between the two calculations at the per-mille level for essentially
the entire range of $\tc$ values studied here.  The reason for this is clear in the
case of the $qq$ and $\bar q \bar q$ channels since they enter for the first time at this order
and only contain collinear singularities.
Moreover, for the $q\bar q$ channel the dominant contribution comes not from $s$-channel diagrams
that are present at LO, but from $t$-channel scattering diagrams that only enter at NLO and have
a similar singularity structure as those for $qq$ and $\bar q \bar q$.  Since the effect of power corrections is so
small we see essentially no gain in using the boosted definition of ${\mathcal T}_1$.

Since the boosted definition performs better, it is clear that we should use it for assessing the performance of
the jettiness calculation.  In order to summarize our findings we will compare with the exact NLO result,
for two cases.  In the first we simply use $\epsilon = 5 \times 10^{-5}$, while in the second we define
$\delta_{NLO}^{fit}$ as the asymptotic fit value indicated in Eq.~(\ref{eq:fitform-nlo1}) for the leading channels and
simply use $\epsilon = 2.5 \times 10^{-5 }$ for the subleading channels.  This comparison is shown
in Table~\ref{tab:h1jnnlojet-nlo}.  We conclude that either choice reproduces the exact result at the
$0.15\%$ level or better.
\begin{table}[!h]
\begin{tabular}{@{}llllllll@{}}
\hline 
Calculation    & ~~~$gg$          & ~~~$qg$               & ~~~$\bar{q}g$      & ~~~$q\bar{q}$        & ~~~$qq$             & ~~~$\bar{q}\bar{q}$   & ~~~total \\ 
\hline
$\epsilon = 5 \times 10^{-5}$~
               & $ 4967 \pm 11 $~~& $  547.3 \pm 1.0 $   & $ 231.5 \pm 0.4 $ & $ -14.65 \pm 0.03 $ & $ -33.92 \pm 0.05$ & $ -6.74 \pm 0.01 $ & $6455 \pm 19$\\ 
$\delta_{NLO}^{fit}$
               & $ 4960 \pm  8 $  & $  547.3 \pm 0.7 $   & $ 231.3 \pm 0.3 $ & $ -14.57 \pm 0.03 $ & $ -33.84 \pm 0.04$ & $ -6.73 \pm 0.01 $ & $6447 \pm  9$\\
Exact          & $ 4960 \pm  2 $~~& $  546.3 \pm 0.4 $  ~~& $ 231.1 \pm 0.1 $~~& $ -14.62 \pm 0.04 $~~& $ -33.94 \pm 0.08$~~& $ -6.73 \pm 0.01 $~~& $6445 \pm  3$\\
\hline
\end{tabular}
\caption{Comparison between NLO coefficients computed by MCFM, both exactly (using dipole subtraction) and
by jettiness slicing (boosted definition of ${\mathcal T}_1$).
Results are shown for $\epsilon = 5 \times 10^{-5}$ in the definition of $\tc$ and for a combination of
fit values ($gg$, $qg$, $\bar q g$) and results for $\epsilon = 2.5 \times 10^{-5}$ ($q\bar q$, $qq$, $\bar q \bar q$),
denoted by $\delta_{NLO}^{fit}$.
Note that the total column includes a factor of two for channels that are not beam-symmetric.}
\label{tab:h1jnnlojet-nlo}
\end{table}

\subsection{Comparison of NNLO calculation}
\begin{figure}
\begin{center}
\includegraphics[width=0.45\textwidth]{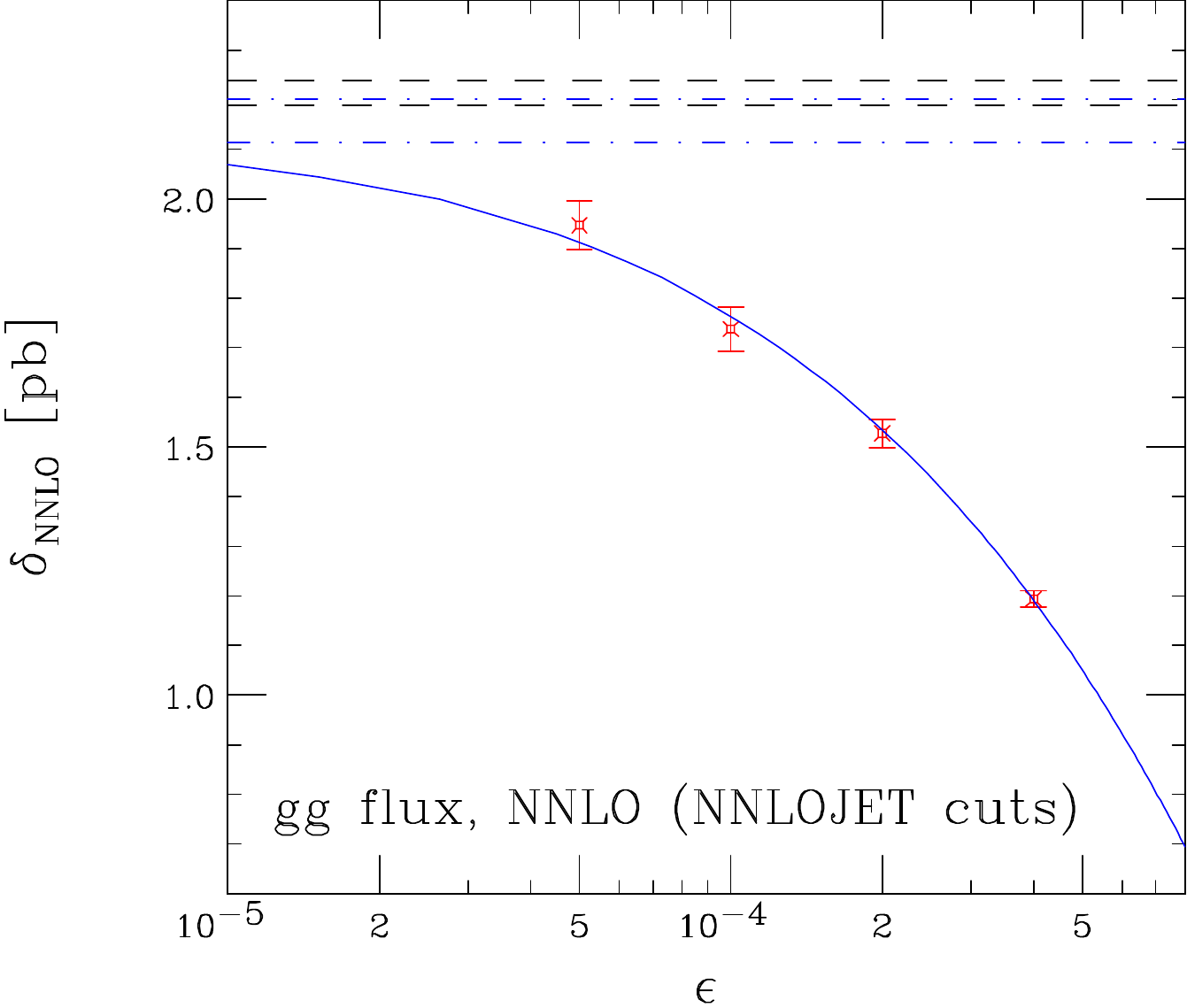}\hspace*{0.5cm}
\includegraphics[width=0.45\textwidth]{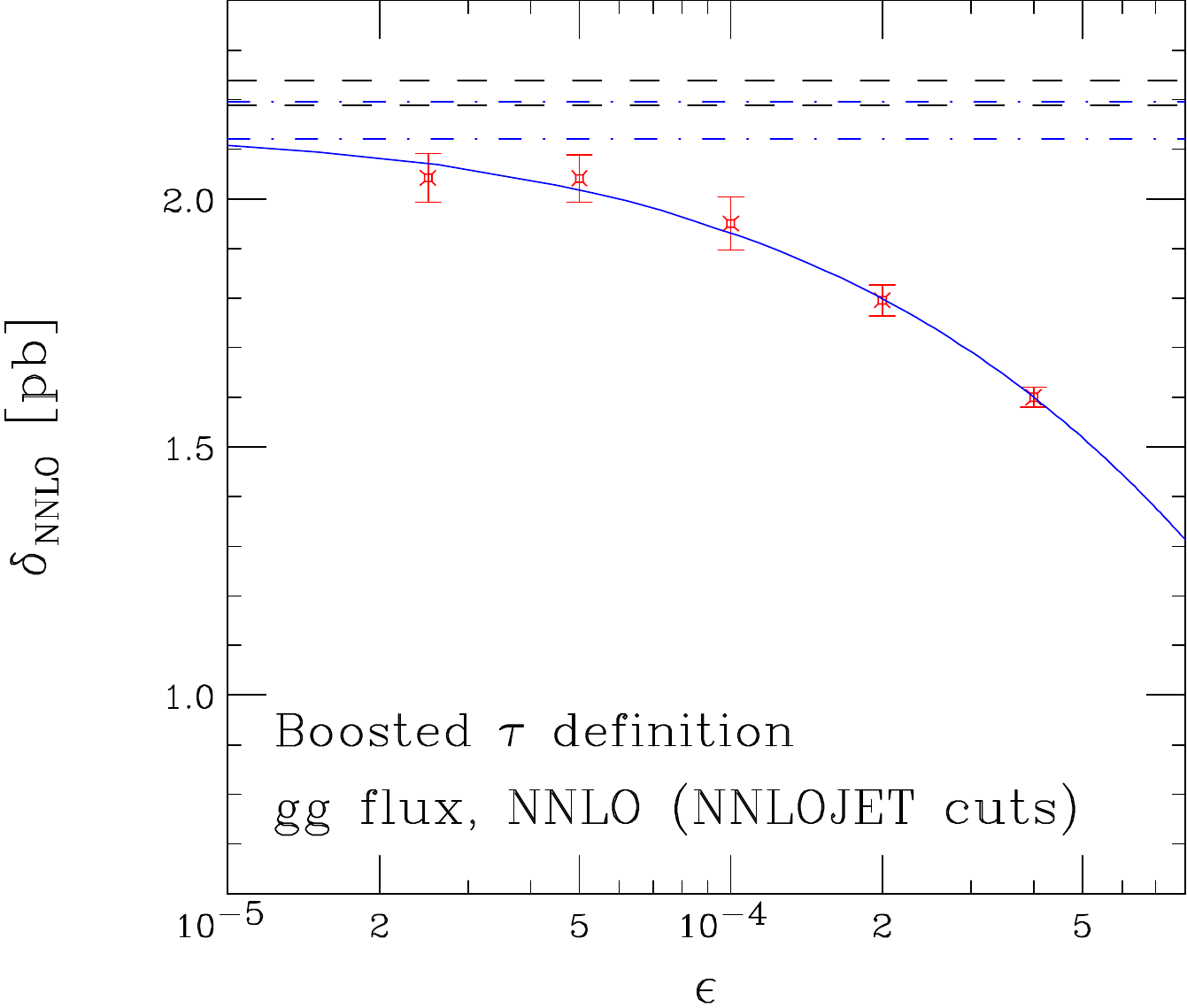} \\ \vspace*{0.45cm}
\includegraphics[width=0.45\textwidth]{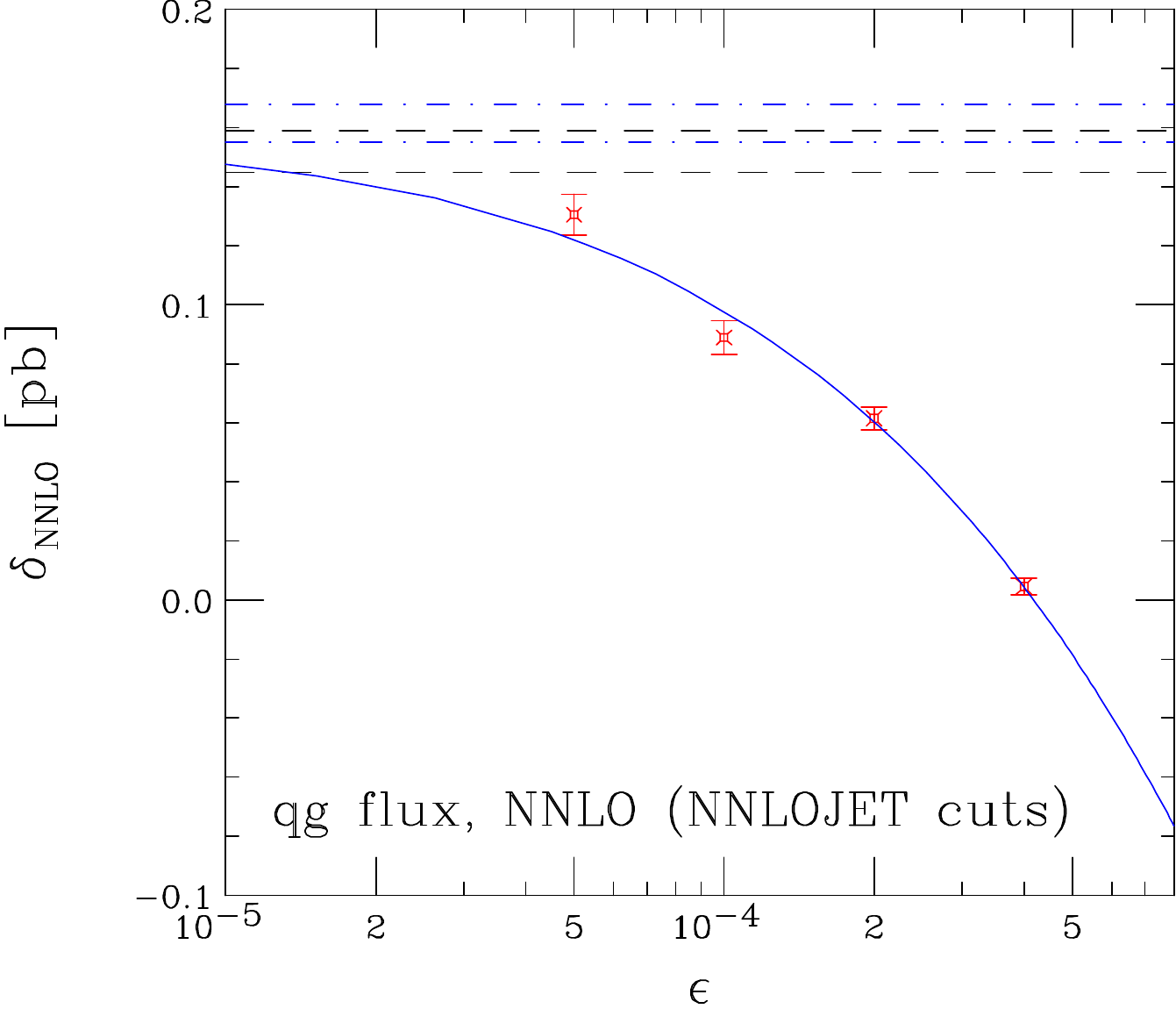}\hspace*{0.5cm}
\includegraphics[width=0.45\textwidth]{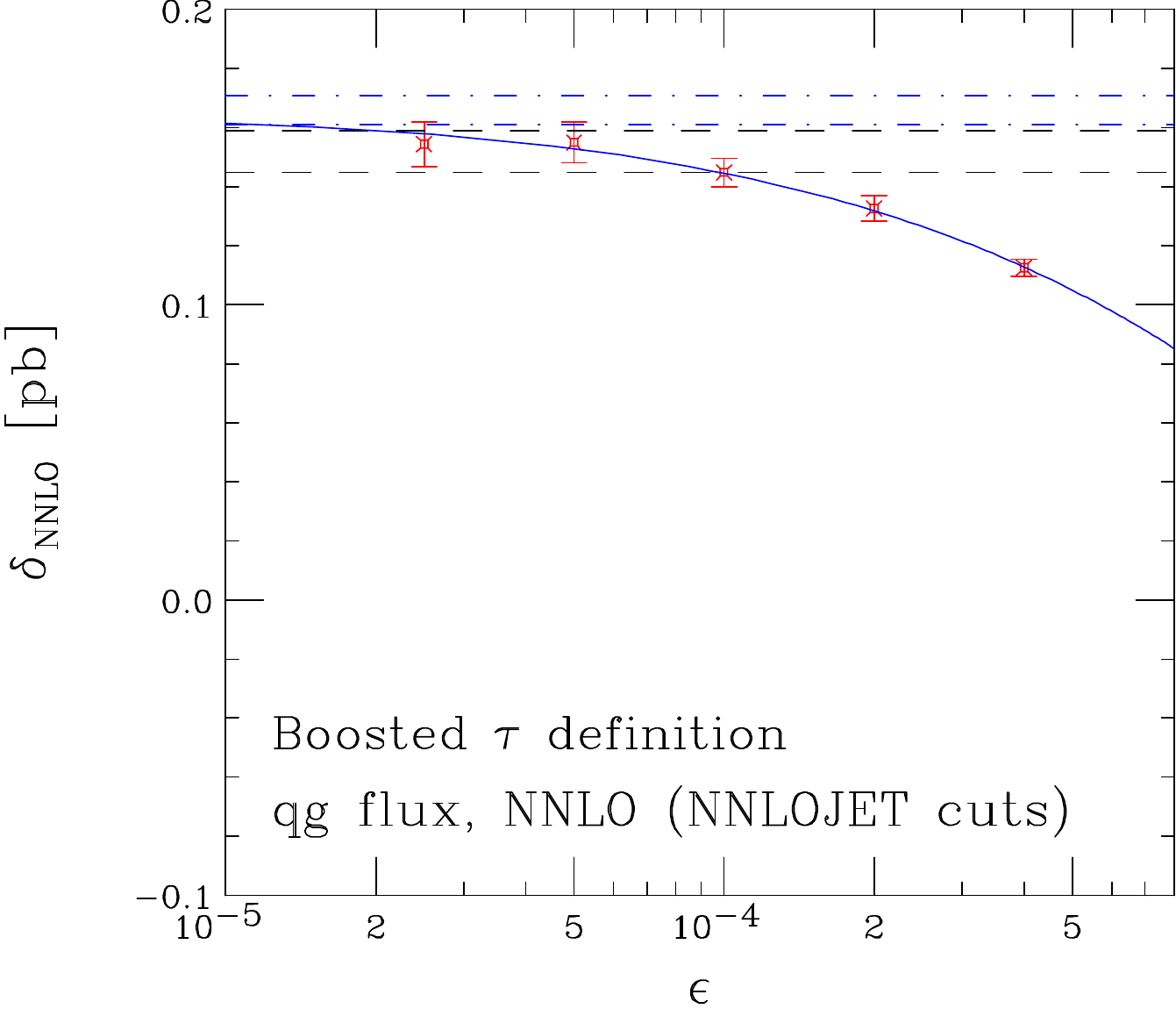} \\ \vspace*{0.45cm}
\includegraphics[width=0.45\textwidth]{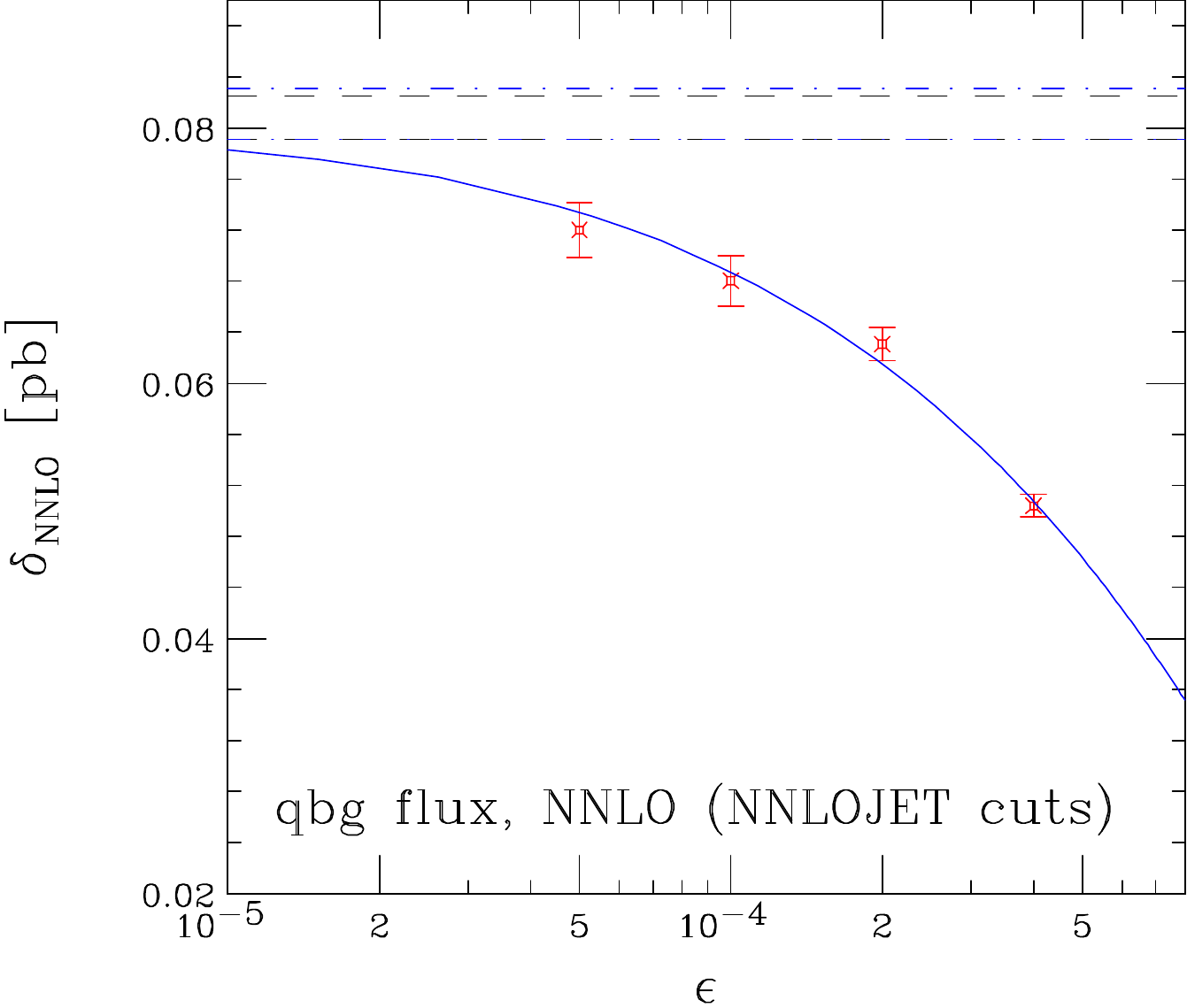}\hspace*{0.5cm}
\includegraphics[width=0.45\textwidth]{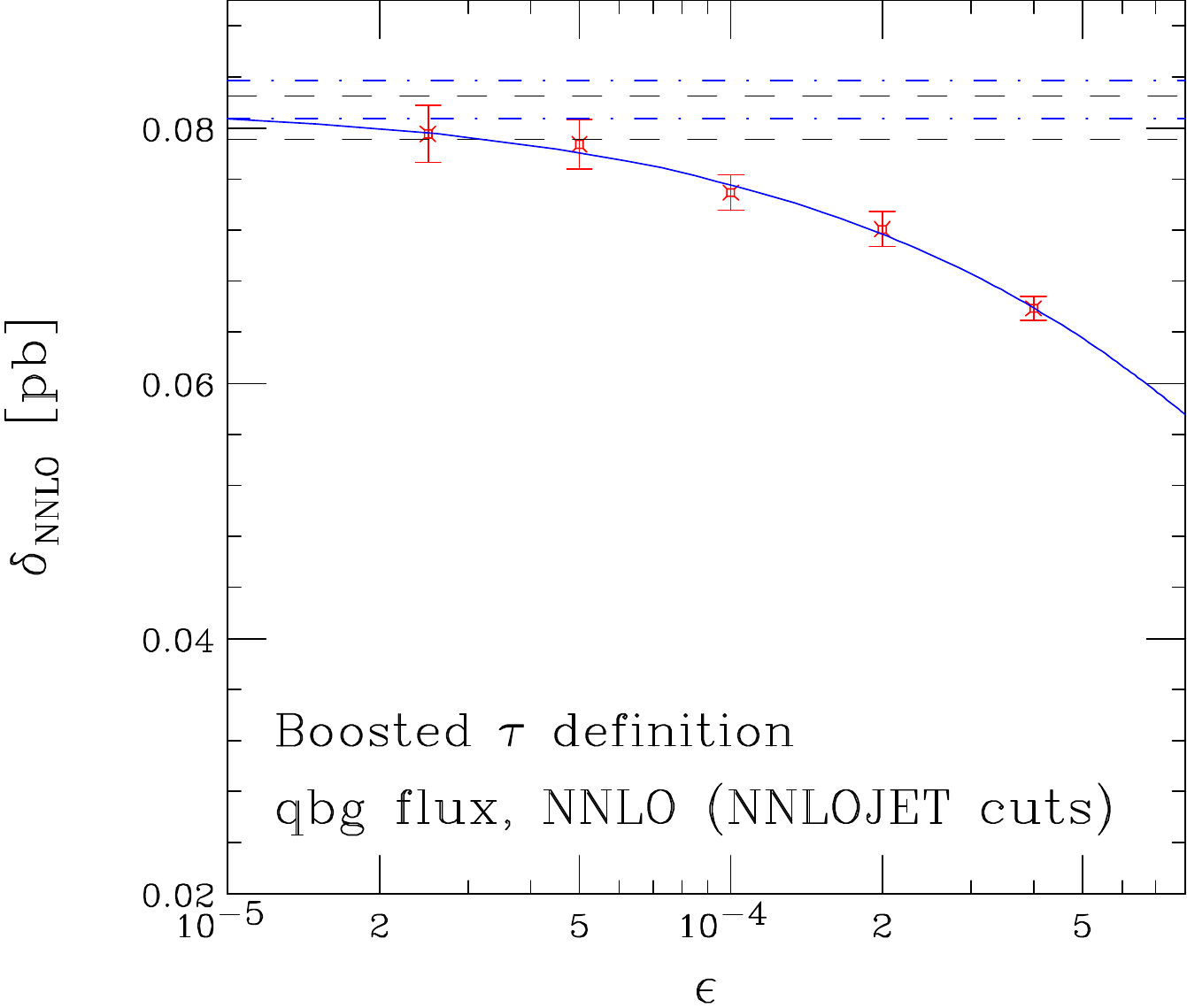}
\end{center}
\caption{$\tau$-dependence of NNLO coefficients for the $gg$, $qg$ and $\bar q g$ partonic channels, in the NNLOJET setup.
The plots on the left show the result when ${\mathcal T}_1$ is computed in the hadronic c.o.m. and the ones on the right
indicate the corresponding result when evaluating this quantity in the boosted frame.
The (blue) solid lines correspond to the fit form in Eq.~(\ref{eq:fitform-nnlo1}), with the dot-dashed lines representing
the errors on the asymptotic value of the fit. The NNLOJET result, including its associated uncertainty,
is shown as the band enclosed by the black dashed lines.}
\label{fig:taudep-nnlojet-leading}
\end{figure}

We now turn to an examination of the NNLO calculation, for which we perform a similar $\tc$-dependence study.
As before, we first inspect the performance of the calculation in the leading partonic channels that are
subject to the largest power corrections, using both versions of ${\mathcal T}_1$.
The results are shown in Fig.~\ref{fig:taudep-nnlojet-leading}, which indicates again that using the boosted
definition of ${\mathcal T}_1$ results in a less dramatic approach to the asymptotic result.
In contrast to the case at NLO, but as anticipated from
the stronger power corrections that are present at this order, the dependence on $\tc$ is quite pronounced.
The region in which the power corrections are under control
is much reduced, even when using the boosted definition of ${\mathcal T}_1$.
The results only begin to become independent of $\tc$, at around the $5\%$ level, for
$\epsilon = 10^{-4}$ or smaller.  The figures also indicates the results of a fit to the data using the
expected form of the power corrections at this order, which takes the form,
\beq
\delta_{NNLO}^{\{gg, qg, \bar qg\}}(\epsilon) = \delta_{NNLO}^{fit} + c_0 \, \epsilon \log^3(\epsilon) + \ldots
\label{eq:fitform-nnlo1}
\eeq
This leading behavior is sufficient for the boosted definition but we observe that for ${\mathcal T}_1$
defined in the hadronic c.o.m. it may be more appropriate to include an additional subleading
$\epsilon \log^2(\epsilon)$ term.  Since the boosted definition is clearly superior, and well-described by
the leading coefficient alone, we do not investigate this further.   For both definitions we see that the
fit value is in very good agreement with the NNLOJET result.

\begin{figure}
\begin{center}
\includegraphics[width=0.45\textwidth]{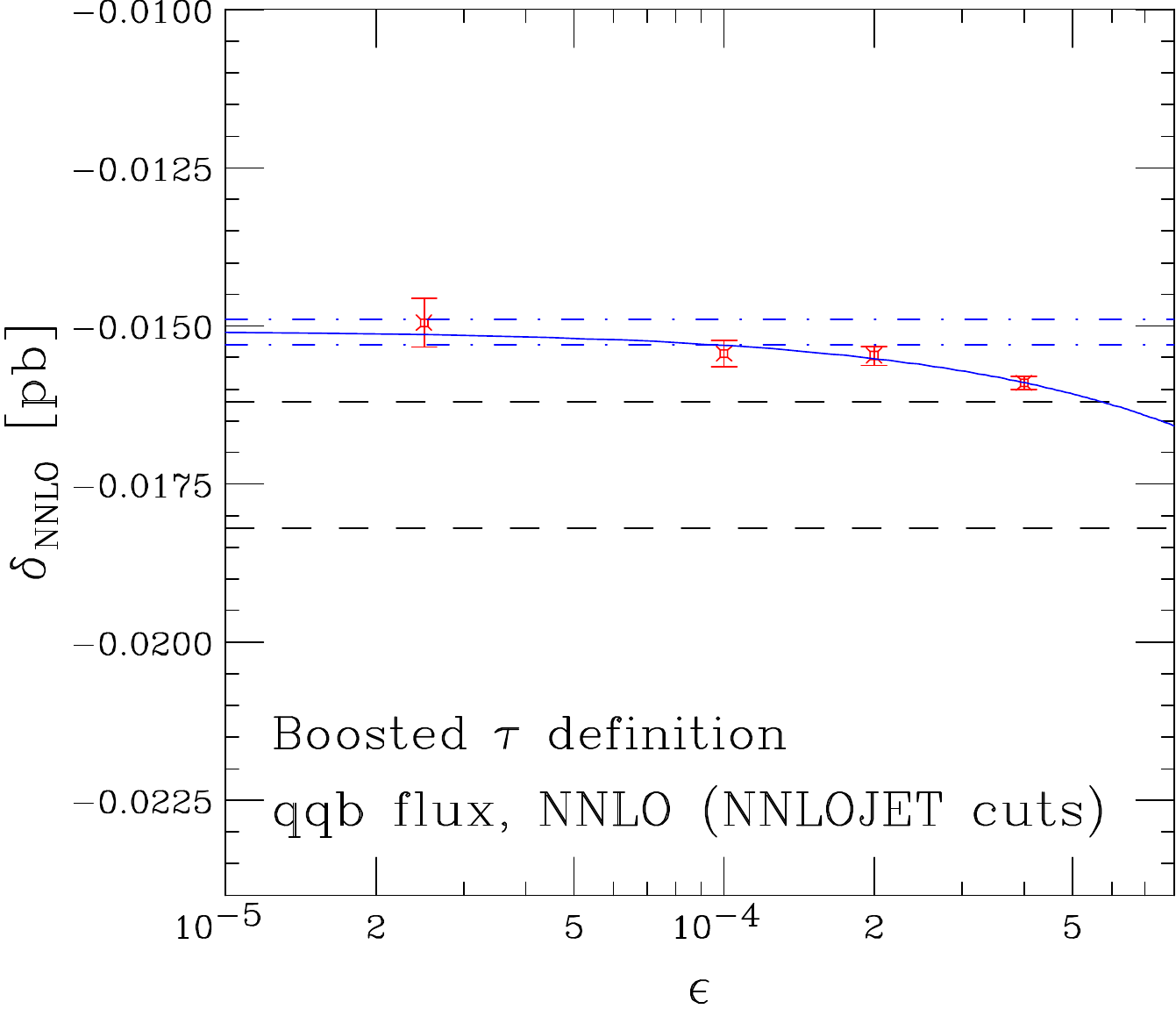} \\ \vspace*{0.5cm}
\includegraphics[width=0.45\textwidth]{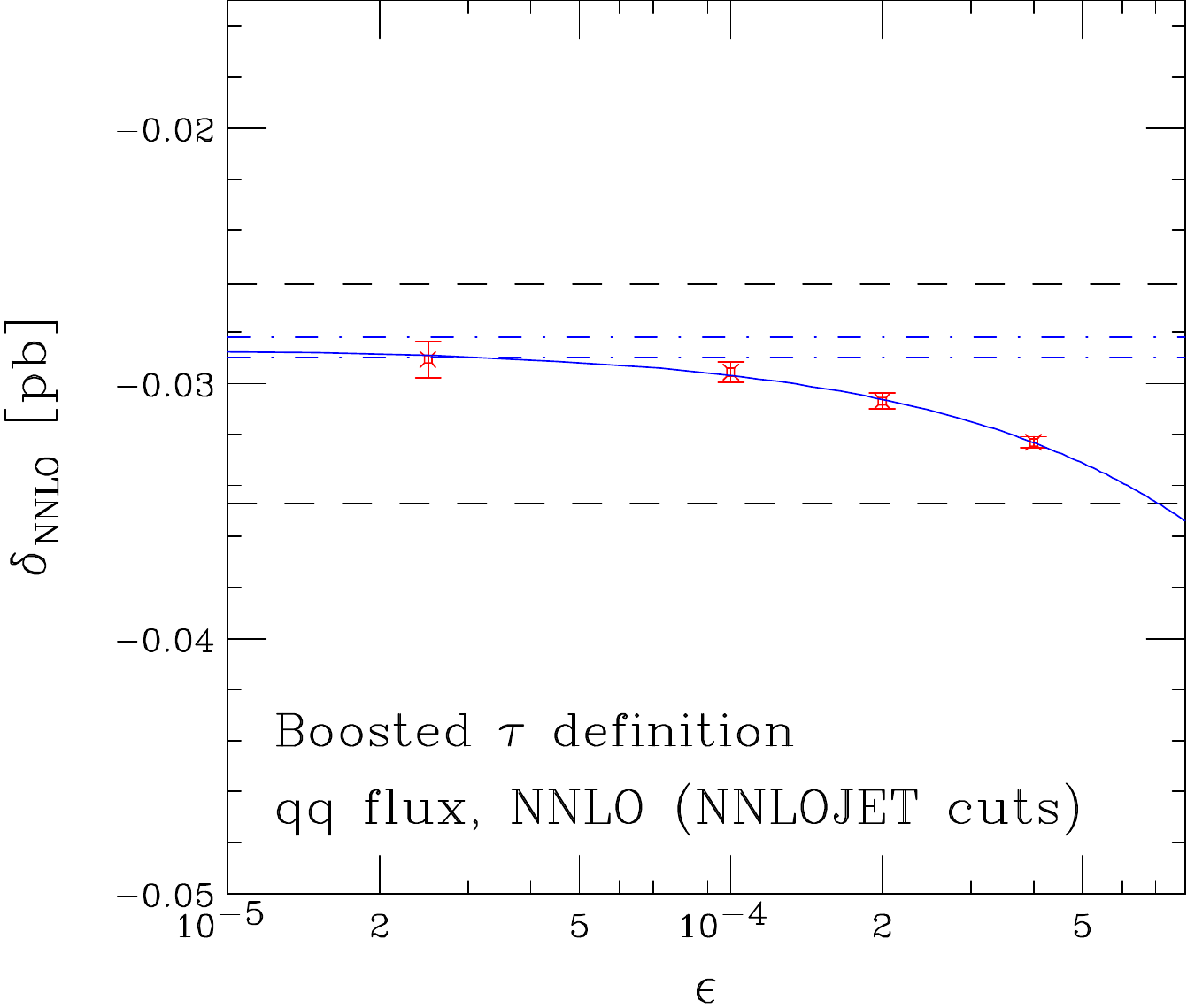} \\ \vspace*{0.5cm}
\includegraphics[width=0.45\textwidth]{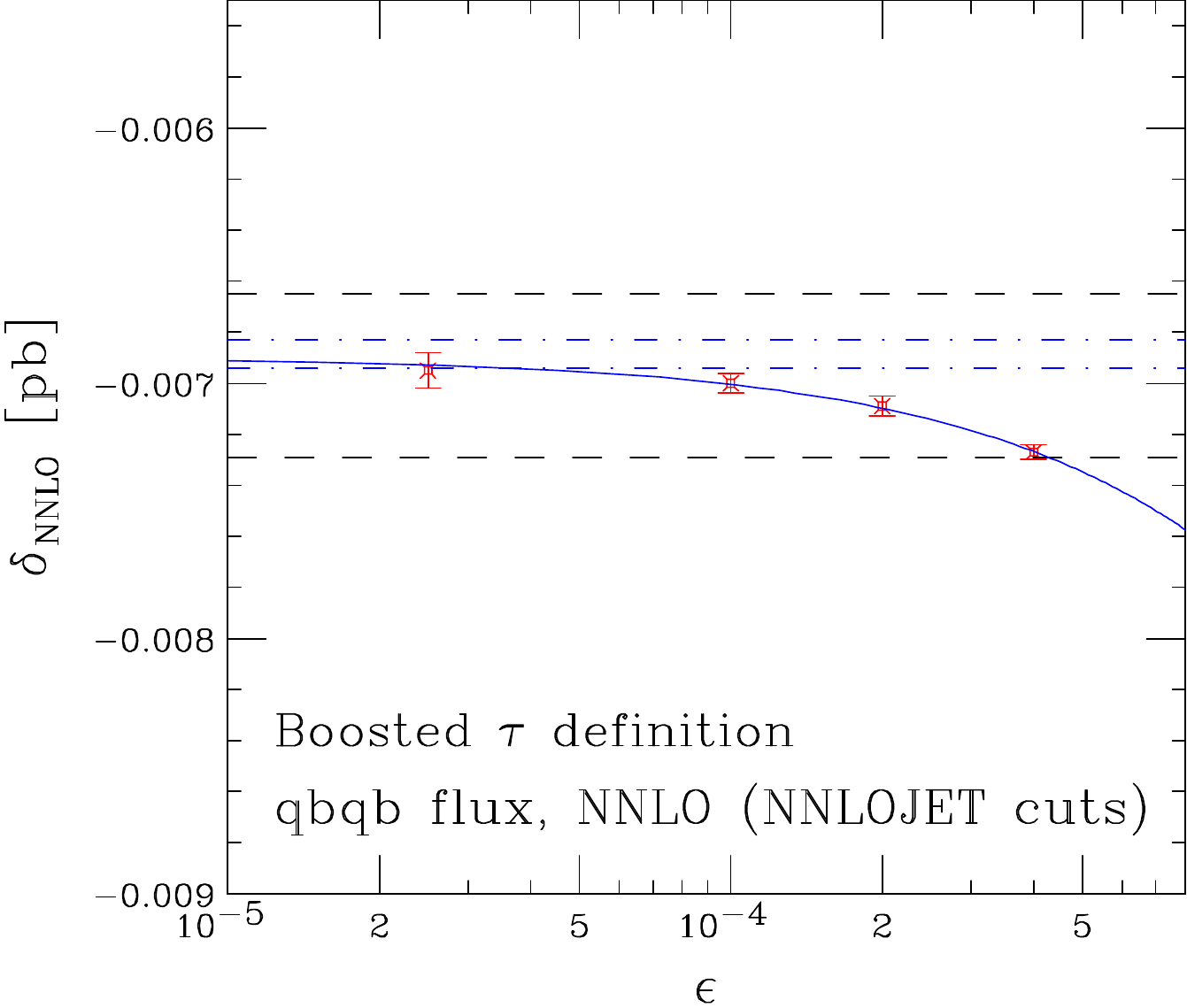}
\end{center}
\caption{$\tau$-dependence of NNLO coefficients for the $q \bar q$, $qq$ and $\bar q \bar q$ partonic channels, in the NNLOJET setup,
using ${\mathcal T}_1$ evaluated in the boosted frame.
The (blue) solid lines correspond to the fit form in Eq.~(\ref{eq:fitform-nnlo2}), with the dot-dashed lines representing
the errors on the asymptotic value of the fit.   The NNLOJET result, including its associated uncertainty,
is shown as the band enclosed by the black dashed lines.}
\label{fig:taudep-nnlojet-subleading}
\end{figure}

A similar study of the subleading channels is shown
in Fig.~\ref{fig:taudep-nnlojet-subleading}, although in this case we choose
to show only the results obtained using the boosted definition of
${\mathcal T}_1$ since it is clear that the power corrections are small.
In all cases there is very little dependence on $\tc$ and the resulting NNLO corrections are
in good agreement with those from NNLOJET, apart from the $q \bar q$ channel that is slightly
outside the error estimate.  However, we note that the NNLOJET calculation with which we compare
did not isolate individual channels and is therefore heavily focussed on the dominant $gg$ and $qg$
channels.  As explained in Ref.~\cite{Chen:2018pzu}, these subleading channels are more
sensitive to numerical fluctuations at larger values of $x$, which may explain the relatively poorer agreement
observed in Fig.~\ref{fig:taudep-nnlojet-subleading}.
For the $1$-jettiness calculation in MCFM we have indicated a fit to the power corrections using a
form that reflects their weaker role in these channels,
\beq
\delta_{NNLO}^{\{q\bar q, qq, \bar q\bar q\}}(\epsilon) = \delta_{NNLO}^{fit} + c_0 \, \epsilon \log(\epsilon) + \ldots
\label{eq:fitform-nnlo2}
\eeq
However we note that, although the $\tc$ dependence is milder for the subleading channels,
the $\tc$ dependence of the total NNLO correction --- and hence the effectiveness of this method ---  is governed
by the behavior of the leading channels.

The final comparison between MCFM and NNLOJET, including also the results from the fits,
is shown in Table~\ref{tab:h1jnnlojet-final}.  Note that we also include, separately and for reference,
the contribution from the Wilson coefficient correction that enters at NNLO.  Note that this contribution
may be computed exactly (without any $\tc$ dependence) since it is simply related to the NLO coefficient.
Since the $\tc$-dependence is stronger at NNLO we use $\epsilon = 2.5 \times 10^{-5}$ as the point
at which we compare our non-fitted results. 
We conclude that this value reproduces the NNLOJET result to within about $5-10\%$ for all channels,
with a significant improvement in the agreement -- especially for the leading $gg$ channel -- when using
the fitted asymptotic result.  

\begin{table}[!h]
\begin{tabular}{@{}llllllll@{}}
\hline
Calculation    & ~~~$gg$~~          & ~~~$qg$~~           & ~~~$\bar{q}g$~~    & ~~~$q\bar{q}$~~       & ~~~$qq$~~           & ~~~$\bar{q}\bar{q}$~~& ~~~total\\ 
\hline
NNLO Wilson~~  & $  879 \pm  2 $ & $ 93.4 \pm  0.4$~~& $ 40.2 \pm 0.2$  & $ -3.0 \pm 0.0$ ~& $ -6.61 \pm 0.0$    & $ -1.33 \pm 0.01$  & $1132 \pm 3$ \\ 
$\epsilon = 2.5 \times 10^{-5}$~
               & $ 2043 \pm 49 $ & $  154 \pm 7$   &   $ 79.5 \pm 2.3 $ & $ -15.0 \pm 0.4 $  & $ -29.1 \pm 0.7 $ & $ -6.95 \pm 0.07$  & $2444 \pm 69$ \\ 
$\delta_{NNLO}^{fit}$
               & $ 2159 \pm 37 $ & $  166 \pm 5$    &  $ 82.2 \pm 1.5 $ & $ -15.1 \pm 0.2 $  & $ -28.6 \pm 0.4 $ & $ -6.90 \pm 0.04$  & $2590 \pm 51$ \\ 
NNLOJET        & $ 2213 \pm 25$~~& $  152 \pm 7$ ~~&   $ 80.8 \pm 1.7 $~~&$ -17.2 \pm 1.0$ ~~& $ -30.6 \pm 4.1$~~& $ -6.97 \pm 0.32$~~& $2607 \pm 49$ \\ 
\hline
\end{tabular}
\caption{Comparison between MCFM and NNLOJET results for the NNLO coefficient $\delta_{NNLO}$,
defined in Eq.~\ref{eq:orders}, in the YR4 setup detailed in the text.
We also show separately the NNLO Wilson coefficient contribution to $\delta_{NNLO}$.
Results are shown for the boosted definition of ${\mathcal T}_1$, for 
$\epsilon = 2.5 \times 10^{-5}$ and also for the fit values ($\delta_{NNLO}^{fit}$).
Note that the total column includes a factor of two for channels that are not beam-symmetric
and uncertainties on individual channels are combined linearly in the total.}
\label{tab:h1jnnlojet-final}
\end{table}

It is useful to perform a cross-check that also tests the scale-dependence of the full result.  For this we employ
a simple 2-point variation in which both renormalization and factorization scales vary by a factor of two together about the
central choice.  At the preceding orders in perturbation theory we find,
\begin{equation}
\sigma_{LO}(\mathrm{MCFM}) = 7.66 ^{+2.92} _{-1.98} \,\text{pb} \,,
\end{equation}
and,
\begin{equation}
\sigma_{NLO}(\mathrm{MCFM}) = 14.12 ^{+2.83} _{-2.45} \,\text{pb} \,,
\end{equation}
which are in complete agreement with the corresponding results from NNLOJET.
At NNLO we first examine the non-fitted result and find,
\begin{equation}
\sigma_{NNLO}(\mathrm{MCFM}, \epsilon = 2. 5 \times 10^{-5}) = 16.56 \pm 0.07 \, ^{+1.03} _{-1.52} \,\text{pb} \,,
\end{equation}
where the error from the Monte Carlo calculation is shown first, and the scale uncertainty is indicated
by the sub- and super-scripts.  This is to be compared with the corresponding result from NNLOJET,
\begin{equation}
\sigma_{NNLO}(\mathrm{NNLOJET}) = 16.73 \pm 0.05 \, ^{+1.00} _{-1.51} \,\text{pb} \,.
\end{equation}
We see that, since the NNLO corrections are so large, the difference between
the total NNLO result computed with NNLOJET and MCFM is at the $1\%$ level and outside the
(combined) $0.5\%$ Monte Carlo errors. Although this difference does lie well within the residual
NNLO scale uncertainty, the fact
that agreement is only at the percent level potentially limits the range and power of the phenomenology
that may be performed with this result.  However, we note that the use of the asymptotic fits
for the central result yields excellent agreement,
\begin{equation}
\sigma_{NNLO}(\mathrm{MCFM, fit}) = 16.71 \pm 0.05 \, ^{+1.03} _{-1.52} \,\text{pb} \,.
\end{equation}

\begin{figure}
\begin{center}
\includegraphics[width=0.9\textwidth]{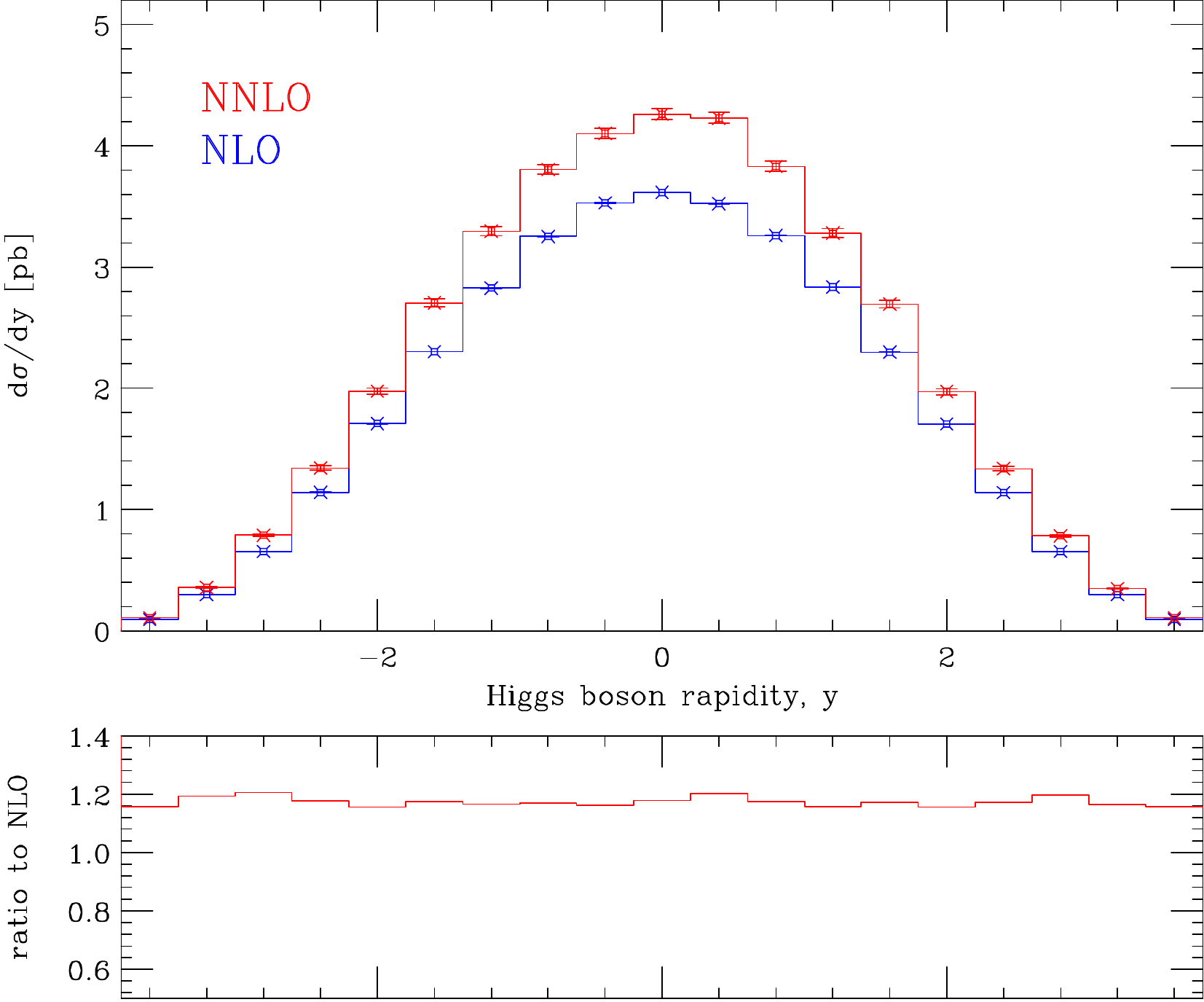}
\end{center}
\caption{The rapidity distribution of the Higgs boson computed at NLO and NNLO using MCFM,
in the NNLOJET setup.
The NNLO coefficient is calculated using $\epsilon = 2. 5 \times 10^{-5}$ in the 
boosted definition of ${\mathcal T}_1$.  The lower panel shows the ratio of the NNLO and
NLO results.}
\label{fig:hrap}
\end{figure}
We conclude this section by examining the calculation of a more differential quantity, the rapidity spectrum of the
Higgs boson.  We show the NLO and NNLO predictions for this observable in Fig.~\ref{fig:hrap},
where the NNLO coefficient is calculated using $\epsilon = 2. 5 \times 10^{-5}$ in the 
boosted definition of ${\mathcal T}_1$.  The effect
of the NNLO corrections is approximately constant in rapidity, with an overall impact that is
in excellent agreement with NNLOJET (c.f. Fig.~24 of Ref.~\cite{deFlorian:2016spz}).


\section{Comparison with BCMPS}
\label{sec:bcmps}

We now turn to a detailed comparison with results obtained using the calculation of
Boughezal, Caola, Melnikov, Petriello and Schulze (BCMPS)~\cite{Boughezal:2015dra}.
Apart from being a cross-check with a different calculation, this comparison provides additional insight
since the setup is slightly different.\footnote{
We thank Fabrizio Caola for providing detailed information
on the calculation used in Ref.~\cite{Boughezal:2015dra} that is used for this comparison.}
The setup for the comparison is as follows:
\begin{eqnarray}
{\rm LHC},~\sqrt{s} = 13~{\rm TeV}, && \quad \mu_R = \mu_F = m_H = 125~{\rm GeV}, \nonumber \\
p_T^{\rm jet} > 20~{\rm GeV}, && \quad {\rm anti-}k_T~{\rm algorithm},~\Delta R = 0.4 \\
{\rm PDF~set:} && \quad {\tt PDF4LHC15\_nnlo\_mc} \nonumber
\end{eqnarray}
In addition, in the calculation of Ref.~\cite{Boughezal:2015dra} NNLO corrections to the
4-quark channels, that first enter the calculation at NLO, are not included.
The essential difference with respect to the previous calculation is the slight reduction in the
jet $p_T$ cut (from $30$ to $20$~GeV), which one expects to render the jettiness calculation
more difficult to perform since the power corrections should be larger for the same value of $\tc$.

As before, we examine the NNLO coefficient alone and separated into partonic channels.
In this case the BCMPS calculation can be easily broken down into three contributions with which we
can compare: $gg$, $qg+\bar qg$ and $q\bar q+qq+\bar q\bar q$, where factors of two to include
all beam-crossings have been included where necessary.  The contributions in these
categories are shown in Table~\ref{tab:h1jcaola}.  As indicated above, in this
calculation the final category -- four-quark channels -- are simply not included at NNLO.  This
is clearly motivated by the size of the contributions at NLO, but is also a check that we can perform
at NNLO with MCFM.

\begin{table}
\begin{tabular}{@{}lllr@{~~\quad}l@{}}
\hline
Contribution  ~~~ & $gg$                & $qg+\bar{q}g$         & $\sum qq$     & Total\\ 
\hline
$\sigma_{LO}$    & $7.957$              & $2.855$               & $0.016$       & $10.828$ \\ 
$\delta_{NLO}$   & $7.422$              & $1.668$               & $-0.139$      &  $8.951$ \\ 
$\delta_{NNLO}$  & $3.408 \pm 0.039$ ~~~& $0.345 \pm 0.008$  ~~~& $0$           &  $3.753$ \\ 
\hline
\end{tabular}
\caption{Cross-sections in picobarns, broken down by channel, using the BCMPS cuts, from
the code used in Ref.~\cite{Boughezal:2015dra}. \label{tab:h1jcaola}}
\end{table}

\begin{figure}
\begin{center}
\includegraphics[width=0.45\textwidth]{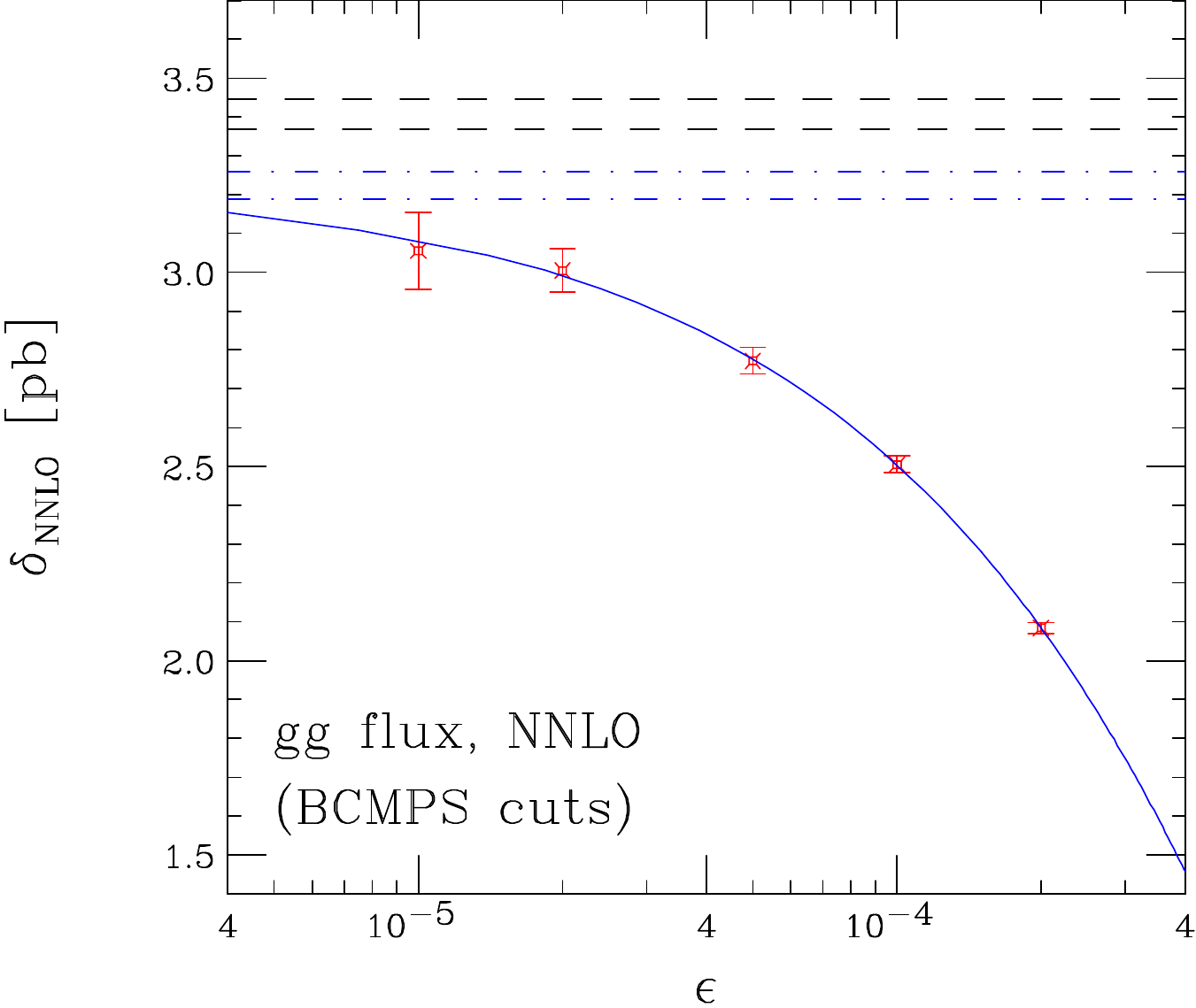}\hspace*{0.5cm}
\includegraphics[width=0.45\textwidth]{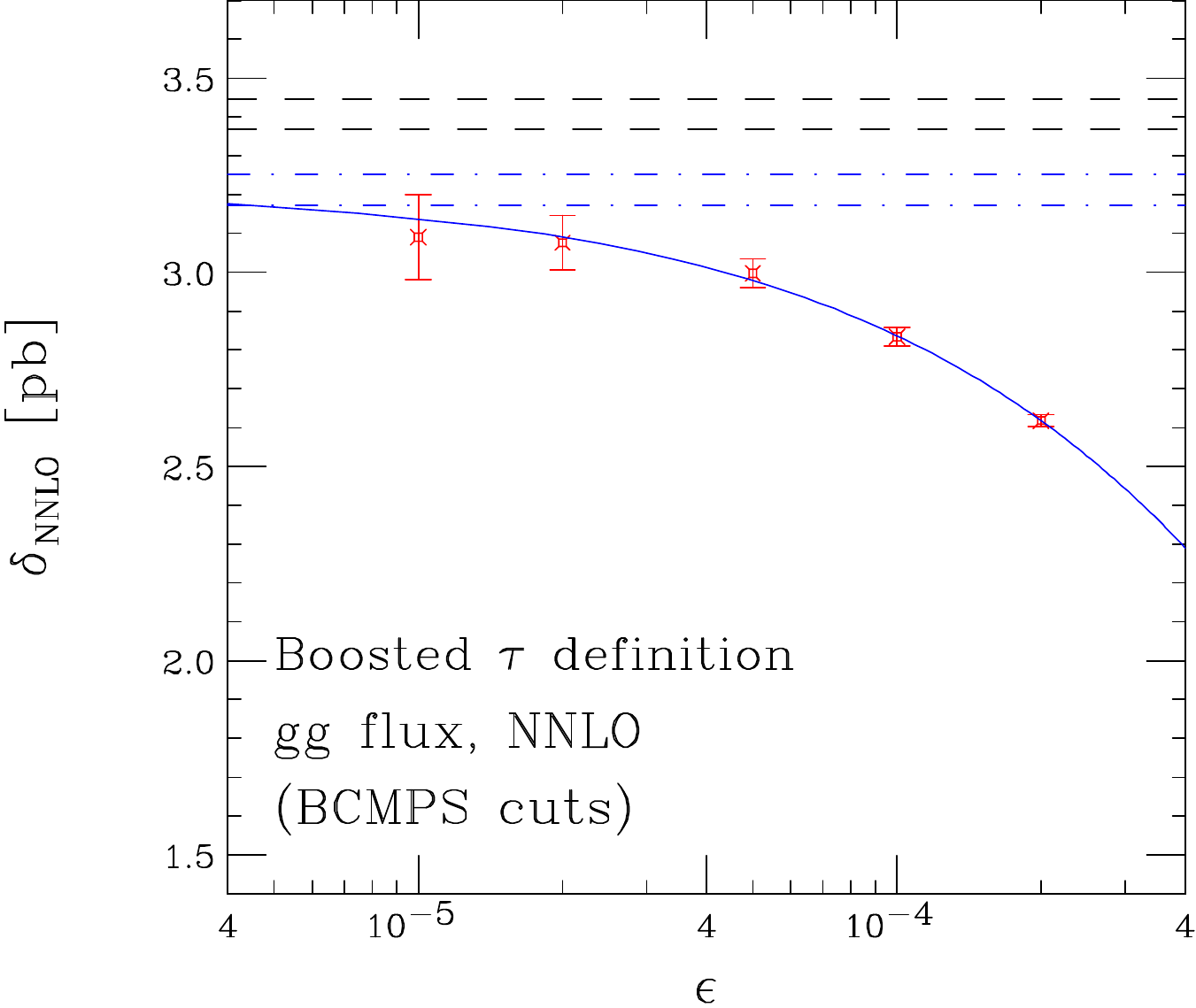}
 \\ \vspace*{0.5cm}
\includegraphics[width=0.45\textwidth]{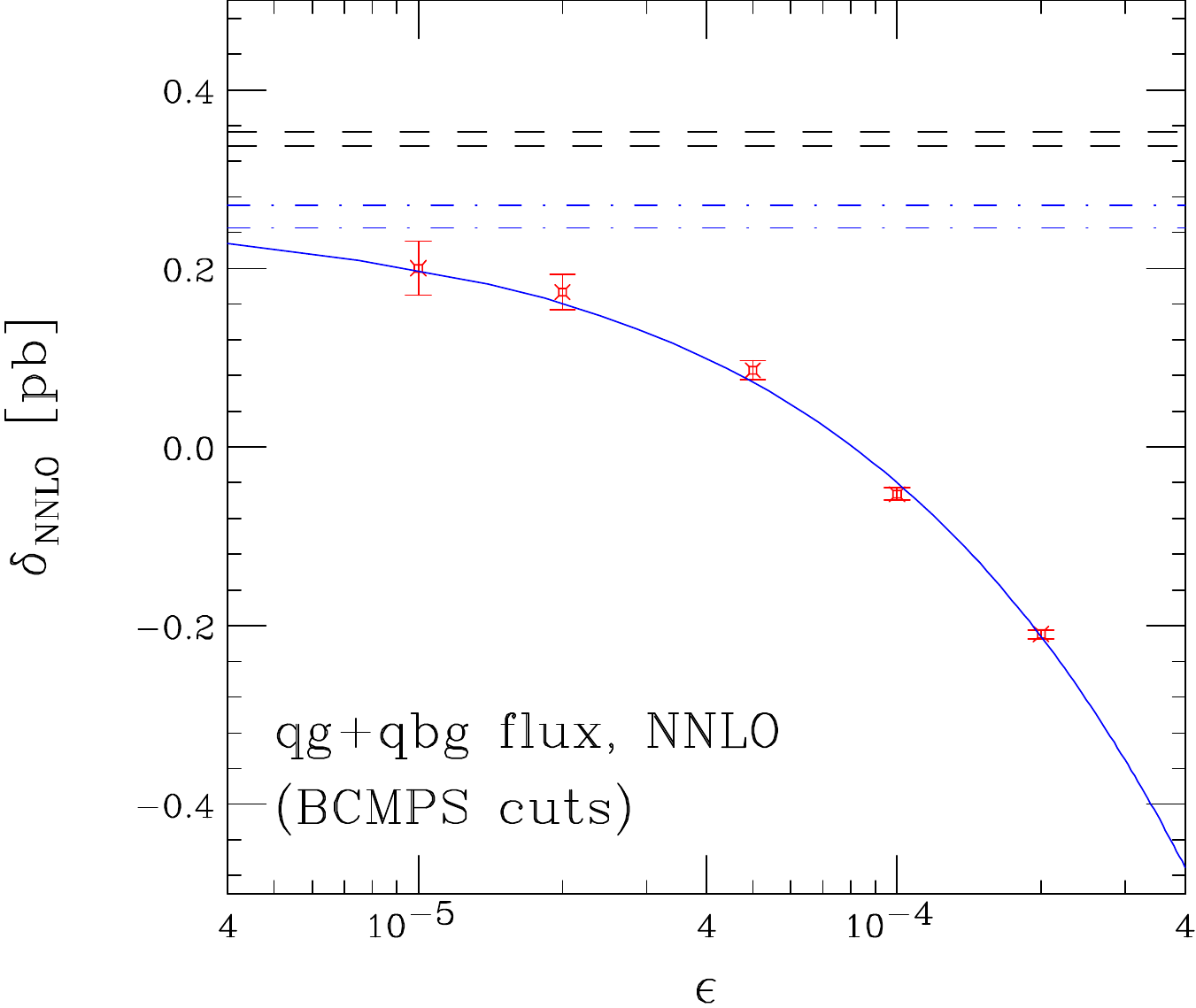}\hspace*{0.5cm}
\includegraphics[width=0.45\textwidth]{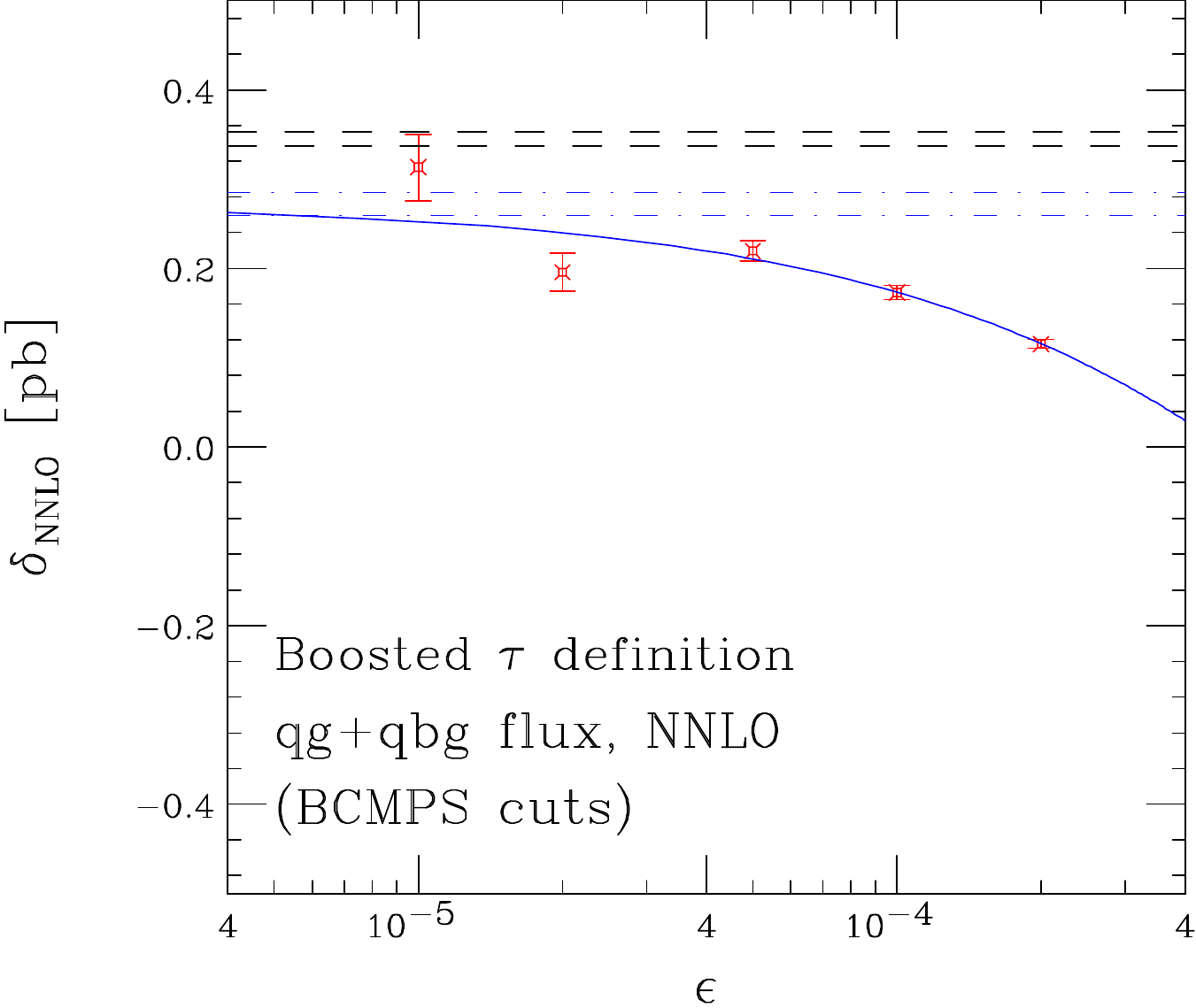}
 \\ \vspace*{0.5cm}
\includegraphics[width=0.45\textwidth]{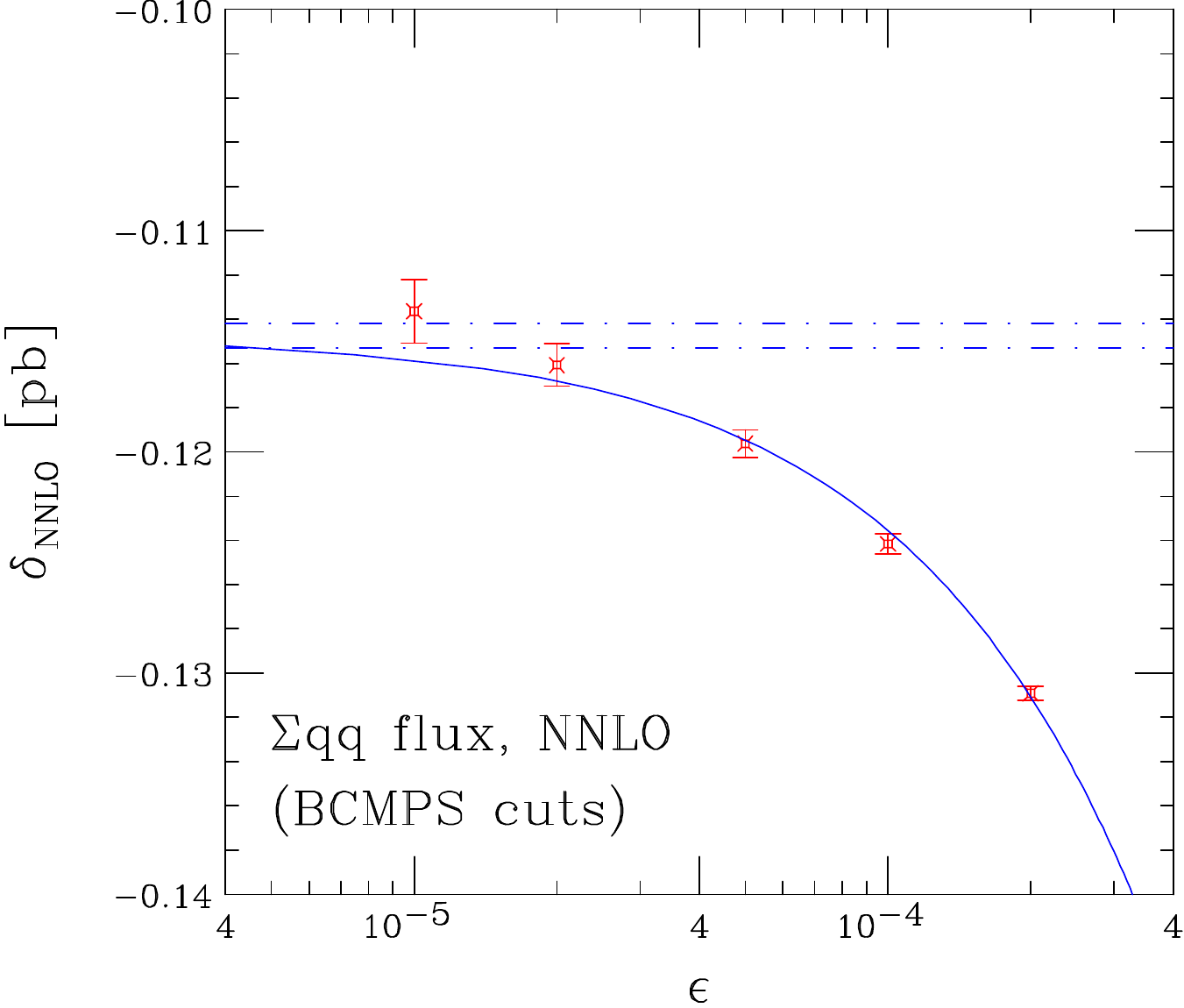}\hspace*{0.5cm}
\includegraphics[width=0.45\textwidth]{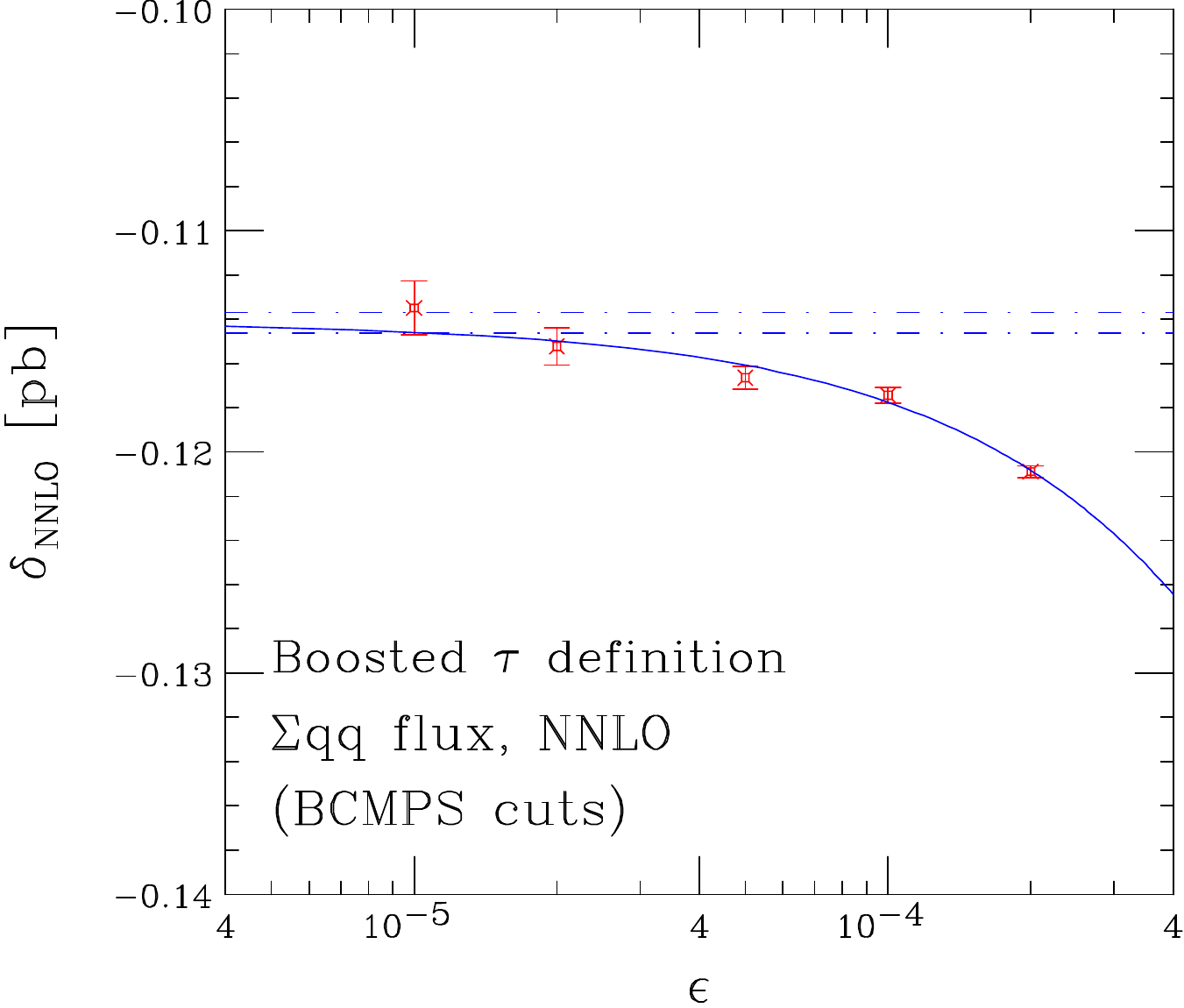}
\end{center}
\caption{$\tau$-dependence of the NNLO coefficient for the $gg$, $qg+\bar qg$ and
four-quark partonic channels using the setup of BCMPS.  The plots on the left-hand side
show the results with ${\mathcal T}_1$ computed in the hadronic c.o.m. while those on
the right are obtained using the boosted definition.
The black dashed lines indicate the BCMPS result, including the uncertainty, and
the bands enclosed by the blue dot-dashed lines show the error on the
asymptotic value obtained from the fitted blue curve.}
\label{fig:fc}
\end{figure}

Results obtained using this setup are shown in Fig.~\ref{fig:fc}, once again for both definitions of ${\mathcal T}_1$.
As before we see that the boosted definition is subject to much weaker power corrections, resulting in a much
quicker approach to the asymptotic result.  For example, at the lowest value of $\tc$ considered here,
corresponding to $\epsilon = 10^{-5}$, the deviation from the asymptotic fit value --  obtained using the
same fit forms as in Section~\ref{sec:nnlojet} -- is around $4\%$ for the $gg$ channel.
We note that this value of $\tc$ is as small as practically possible for
our code, with much lower values becoming sensitive to numerical instability in the evaluation of the
double-real contributions.  
However, we observe that the asymptotic results obtained from this fit to the power corrections indicate somewhat smaller
NNLO corrections to the $gg$ and $qg$ channels than those found by BCMPS.
The asymptotic results for each channel are,
\begin{eqnarray}
&& \delta_{NNLO}^{gg, fit} = 3.213 \pm 0.040 \,\text{pb} \,, \nonumber \\
&& \delta_{NNLO}^{qg+\bar q g, fit} = 0.272 \pm 0.013 \,\text{pb} \,.
\end{eqnarray}
Both results are lower than BCMPS (c.f. Table~\ref{tab:h1jcaola}), by about $6\%$ ($gg$) and $21\%$ ($qg$),
and outside the error bands on the calculations ($1.2\%$ and $5\%$, respectively).  However, we note that
the BCMPS results reported in Table~\ref{tab:h1jcaola} and Fig.~\ref{fig:fc} contain error estimates that
may not be reliable for such a detailed comparison;  they may be underestimated by a factor of around three.
A small difference would still remain for the $qg$ channel even after taking this into account,
which we suspect may be due to our calculation
being unable to go to sufficiently low values of $\tc$ to reliably extract the asymptotic result. 
Taking the original error estimates at face value, the combined effect is a $1.1$\% difference in the total NNLO
cross section -- insufficient to conclusively establish agreement between the calculations in this
region but mostly harmless for phenomenological studies. 

Fig.~\ref{fig:fc} also shows the result of the computation of the correction
to the 4-quark channels ($qq$, $\bar q \bar q$, and $q\bar q$).
Our results verify the size of these corrections at NNLO, with the fitted asymptotic result,
\beq
\delta_{NNLO}^{\sum qq, fit} = -0.114 \pm 0.001 \,\text{pb} \,.
\eeq
This demonstrates that the corrections are of a similar size as the NLO ones, but
are at the level of $0.5\%$ in the total cross-section
and therefore negligible for the purposes of present phenomenology.


\section{Boosted region}
\label{sec:boosted}

We conclude our study with an examination of the performance of the jettiness slicing method in
a region for which it is especially well-suited.  For illustration we consider the calculation
of the cross-section in the boosted region corresponding to a recent CMS analysis searching
for the decay $H\to b\bar b$~\cite{Sirunyan:2017dgc}.  This analysis reconstructs Higgs
boson candidates that satisfy $p_T^H > 450$~GeV, for which the leading theoretical contribution
is a Higgs boson recoiling against a jet of the same transverse momentum.  The cut on $p_T^H$ 
allows a well-defined calculation to be performed at fixed perturbative order, although at higher
orders the cross-section receives contributions from partons of lower momenta.  Nevertheless, at NNLO
such contributions satisfy $p_T > p_T^H/3$, which is still a much stronger constraint than
any of the scenarios studied so far.  We therefore expect the jettiness slicing method to be subject
to much smaller power corrections.  Finally we note that the calculation presented here should not
be compared directly with experimental data since, as is well-known, the effective field theory
used to perform the calculation is not valid in the region $p_T^H > m_t$.  Instead one must take
into account the effect of a finite top-quark mass, for example as in Ref.~\cite{Chen:2016zka},
a procedure that can now be performed using exact results at NLO~\cite{Jones:2018hbb}.  Here
we refrain from such an approach in order to focus instead on the efficacy of the jettiness
method itself.

We modify our parameters only slightly for this study.  We use the same setup as in the previous section,
with the exception that we modify the scale choice in order to take into account the transverse
momentum of the Higgs boson.  We thus use,
\begin{equation}
\mu_R = \mu_F = \sqrt{m_H^2 + \left(p_T^H\right)^2}
\end{equation}
and drop any jet requirement, replacing this with the cut $p_T^H > 450$~GeV.  Here we choose
to quote cross-sections that do not include any pseudo-rapidity cut on the Higgs boson, in contrast
to the CMS analysis~\cite{Sirunyan:2017dgc}.  We note instead that such a cut has almost no
effect on the theoretical calculation, reducing the cross-section by $0.1\%$.
For the jettiness slicing calculation we modify the definition of $\tc$ in order to reflect the
role of the transverse momentum of the Higgs boson, rather than that of the jet, in the definition
of the hardness of the process,
\begin{equation}
\tc = \epsilon \times \sqrt{m_H^2 + \left(p_T^H\right)^2} \,.
\end{equation}

\begin{figure}
\begin{center}
\includegraphics[width=0.45\textwidth]{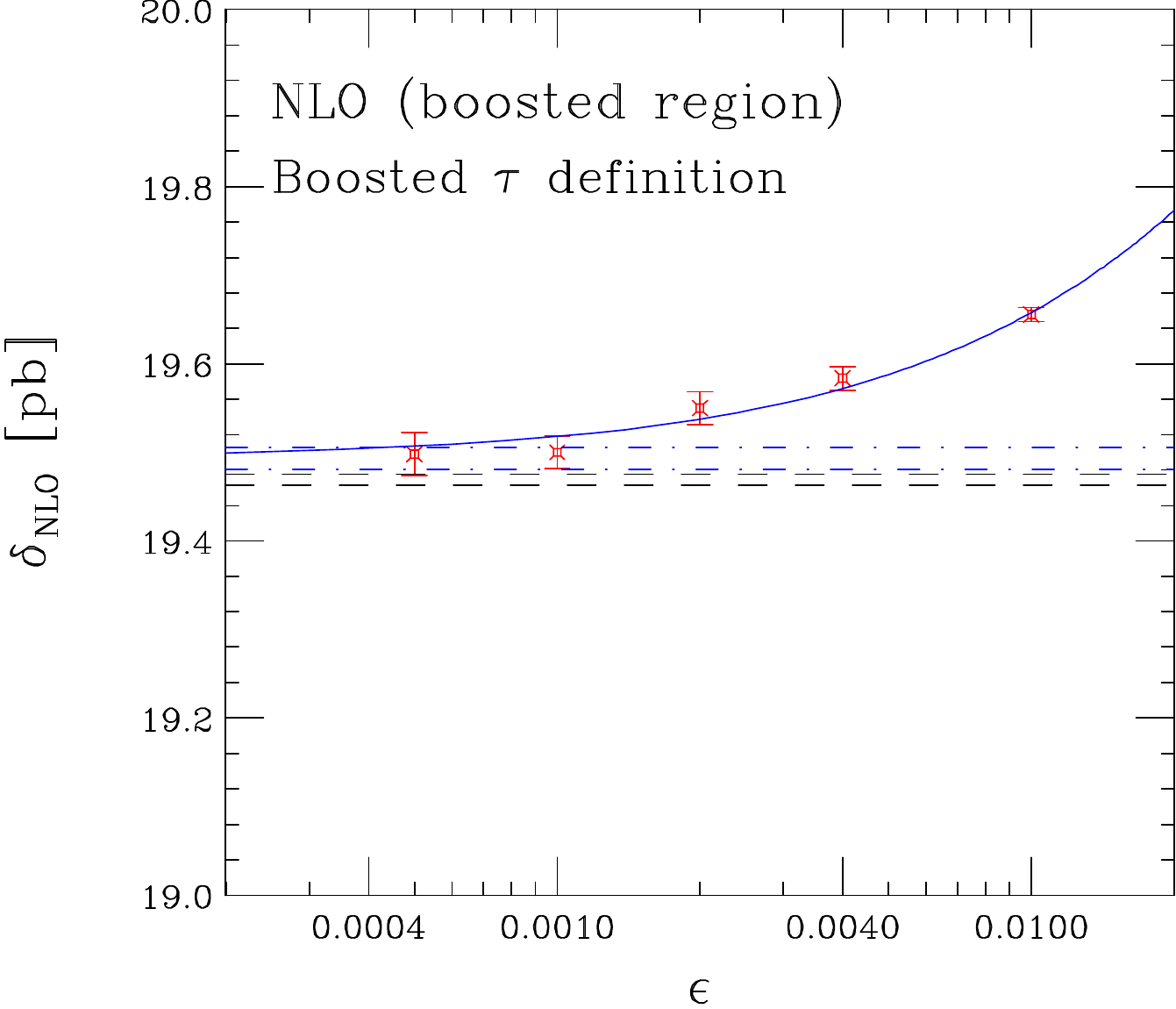}\hspace*{0.5cm}
\includegraphics[width=0.45\textwidth]{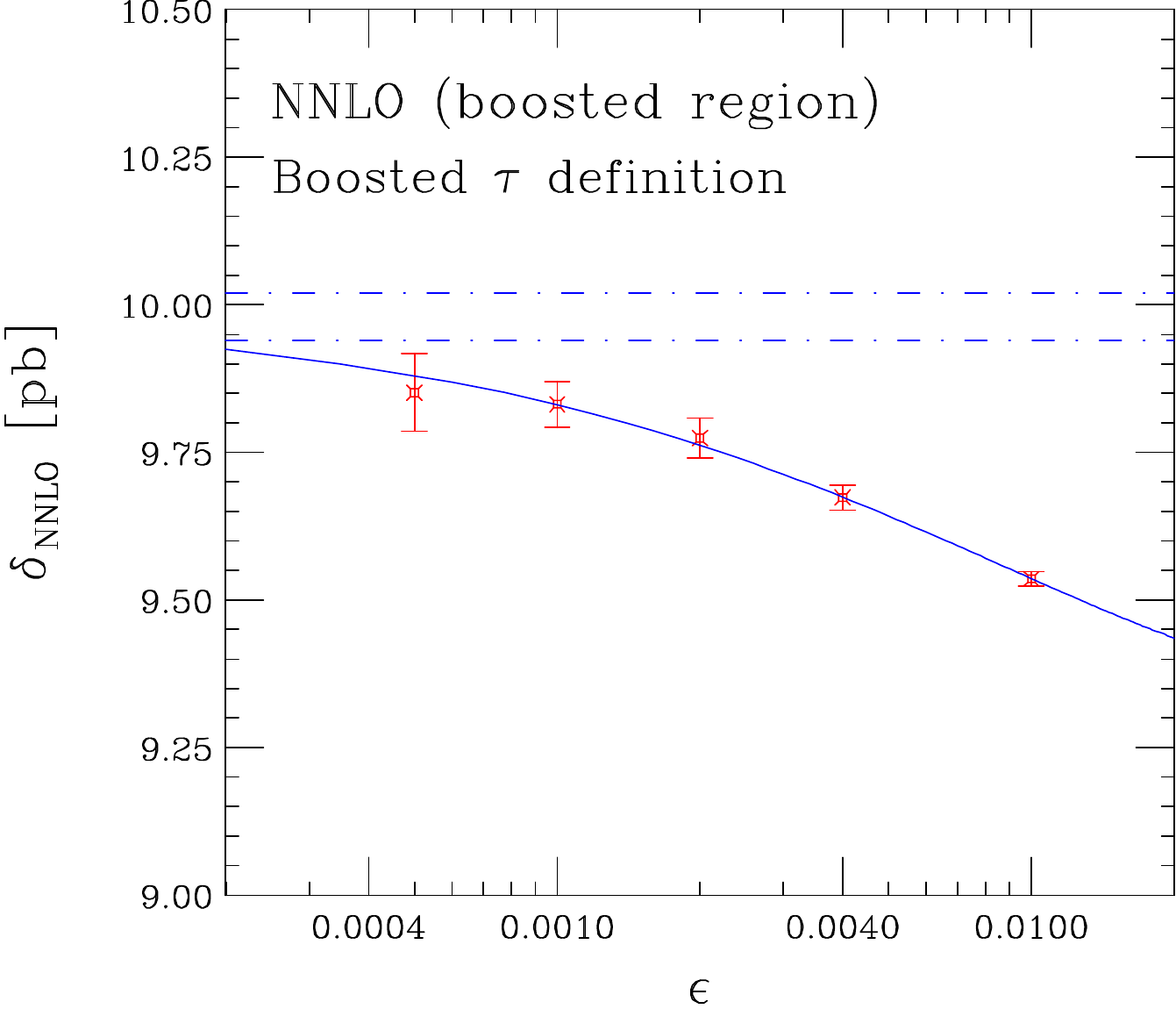} 
\end{center}
\caption{$\tau$-dependence of NLO (left) and NNLO (right) coefficients for Higgs boson
production in the boosted regime, $p_T^H > 450$~GeV.  The (blue) solid lines correspond to the fit forms in
Eqs.~(\ref{eq:fitform-nlo1}) and~(\ref{eq:fitform-nnlo1}), with the dot-dashed lines representing the errors on the
asymptotic value of the fit.  For the NLO coefficient the exact result computed in MCFM using dipole subtraction is
shown as the black dashed line.}
\label{fig:taudepboosth1j}
\end{figure}

The expectation of reduced power corrections in the
boosted region is first confirmed by the results of a study at NLO, shown in
Fig.~\ref{fig:taudepboosth1j}~(left).  In this case the jettiness slicing results agree with
those of the exact calculation at NLO, to within 0.6\%, even for $\epsilon = 4 \times 10^{-3}$.  For comparison,
we observe that a similar level of agreement for the jet cut in Section~\ref{sec:nnlojet} ($p_T^{\text{jet}} > 30$~GeV)
is only obtained for $\epsilon = 2 \times 10^{-4}$ .
From Fig.~\ref{fig:taudepboosth1j}~(right) it is clear that the
calculation of the NNLO coefficient is similarly improved in the boosted region, with the agreement
between the fit result and the point at $\epsilon = 10^{-3}$ already at the $1.5$\% level.  
When combined with the NLO cross-section,
\begin{equation}
\sigma_{NLO}(p_T^H>450~\mbox{GeV}) = 40.67 \,\text{pb} \,,
\end{equation}
we find,
\begin{equation}
\sigma_{NNLO}(p_T^H>450~\mbox{GeV}, \epsilon=10^{-3}) = 50.50 \pm 0.04 \,\text{pb} \,.
\end{equation}
Therefore the difference between this result and the one that would be obtained with the asymptotic fit
is around $0.3\%$, well below the level of phenomenological interest.  We note in passing that the
effect of the NNLO corrections on the boosted cross-section is only slightly larger than at lower
transverse momenta.


\section{Conclusions}
\label{sec:conclusions}

In this paper we have presented a calculation of $H$+jet production at NNLO using the
$N$-jettiness procedure.  This calculation shares many elements with an earlier computation
using the same method~\cite{Boughezal:2015aha}, but differs in the exact implementation.  In
particular, small errors in the above-cut $H+2$~jet NLO calculation have been corrected and
the analysis has been performed at smaller values of the jettiness-slicing parameter, $\tc$.
We have compared results with other calculations available in the
literature~\cite{Chen:2014gva,Chen:2016zka,Bizon:2018foh,Boughezal:2015dra} and found good
agreement.   As anticipated from the jet cuts used for the comparisons, in particular the
relatively low transverse momenta and lack of any rapidity requirement, the $N$-jettiness 
calculations suffers from relatively large power corrections.  These can be ameliorated by
using a definition of $1$-jettiness that accounts for the boost of the Higgs+jet system.
For these comparisons we showed
that it is possible to determine the NNLO coefficient $\delta_{NNLO}$ with an accuracy of around $5\%$ with
reasonable numerical stability, but that substantially better agreement can only be obtained
by fitting out the effect of power corrections.  On the other hand,
since $\delta_{NNLO}/\sigma_{NNLO} \approx 1/6$, an accuracy of $5\%$ in the NNLO
coefficient translates into an error on the total rate, $\sigma_{NNLO}$, of less than $1\%$.
We also showed that requiring a substantially harder jet reduces the effect of power corrections
considerably and renders the method more competitive.  Our calculation demonstrates the 
importance of a dedicated program to compute the effects of power corrections analytically,
as has already been performed for the color-singlet
case~\cite{Moult:2016fqy,Boughezal:2016zws,Moult:2017jsg,Boughezal:2018mvf,Ebert:2018lzn},
in order to improve the effectiveness of the $N$-jettiness method.


\section*{Acknowledgements}
We thank Fabrizio Caola, Xuan Chen and Nigel Glover for their help in
performing the cross-checks that are reported in this paper, and for their insightful
comments on a preliminary draft.  The
numerical calculations reported in this paper were performed using the
Wilson High-Performance Computing Facility at Fermilab and the COSMA
Data Centric system (operated by the Institute for Computational
Cosmology) at Durham University.
The research of JMC is supported by the US DOE under contract DE-AC02-07CH11359.

\bibliography{paper}

\end{document}